\documentclass[a4paper,11pt]{article}

\usepackage[
backend=biber,
style=numeric-comp, 
sorting=none 
]{biblatex}
\usepackage[utf8]{inputenc}
\usepackage[british]{babel}
\usepackage{csquotes}

\usepackage{manfnt}
\usepackage{braket}

\newcommand{\COMMENTOOK}[1]{}

\newcommand{\igor}[1]{{\color{green}IP: #1}}

\usepackage{graphicx}
\usepackage{subcaption} 
\usepackage{color} 

\usepackage{authblk}
\usepackage{hyperref}
\usepackage{amsmath}
\usepackage{amssymb}
\usepackage{amsfonts}
\usepackage{enumitem}
\numberwithin{equation}{section}
\usepackage{ytableau}

\usepackage{cleveref}

\usepackage{lscape} 

\newcommand{\lhs}{{l.h.s.} }

\newcommand{\nzm}{{n.z.m.} }
\newcommand{\wrt}{{w.r.t.} }

\newcommand{\lc}{{lightcone} }

\newcommand{\sN}[1]{ { [#1] } }  

\newcommand{\oh}{ \frac{1}{2} }
\newcommand{\ap}{ {\alpha'} }
\newcommand{\dap}{ {2\alpha'} }

\newcommand{\sdap}{ \sqrt{2\alpha'} }
\newcommand{\shap}{ \sqrt{\frac{\alpha'}{2}} }
\newcommand{\ishap}{ \sqrt{ \frac{ 2 }{ \alpha'} } }
\newcommand{\hap}{ \frac{\alpha'}{2} }

\newcommand{\R}{ \mathbb{R} }
\newcommand{\C}{ \mathbb{C} }

\newcommand{\cH}{{\cal H} }

\newcommand{\cM}{{\cal M} }
\newcommand{\cN}{{\cal N} }

\newcommand{\cS}{{\cal S} }

\newcommand{\cV}{{\cal V} }

\newcommand{\uA}{ {\underline A}}

\newcommand{\uL}{ {\underline L}}
\newcommand{\uk}{ {\underline k}}
\newcommand{\up}{ {\underline p}}

\newcommand{\ux}{ {\underline x}}

\newcommand{\glU}{ U }

\newcommand{\Ulc}[1]{ {U_{\sN #1}} }
\newcommand{\Zlc}[1]{ {Z_{\sN #1}} }

\newcommand{\xt}{{x_a}}
\newcommand{\xr}{{x_b}}
\newcommand{\xs}{{x_c}}

\renewcommand{\xt}{{x_t}}
\renewcommand{\xr}{{x_r}}
\renewcommand{\xs}{{x_s}}

\newcommand{\RR}{{r}}
\renewcommand{\SS}{{s}}
\newcommand{\TT}{{t}}
\newcommand{\UU}{{u}}

\newcommand{\oldI}{{I}}

\newcommand{\TOGGLE}[1]{}

\newcommand{\Alc}{{\alpha_{lc}}}
\newcommand{\uLlc}{{{L_{lc}}}}
\newcommand{\uxlc}{{{x_{lc}}}}
\newcommand{\uplc}{{{p_{lc}}}}
\newcommand{\XX}{{X}}
\newcommand{\ZZ}{{Z}}
\newcommand{\WW}{{W}}
\newcommand{\kp}{{k^+}}
\newcommand{\pp}{{p^+}}
\newcommand{\ukp}{{\underline{k}^+}}

\newcommand{\eq}[1]{
    \begin{align}
        #1
    \end{align}
    }

\newcommand{\del}{\partial}
\newcommand{\eps}{\epsilon}

\begin{document}

\begin{center}
  {\Large \bf DDF amplitudes are lightcone amplitudes
    and the naturalness of Mandelstam maps}
\end{center}

\vskip .6cm
\medskip

\vspace*{4.0ex}

\baselineskip=18pt

\begin{center}

{\large 
\rm  Dripto Biswas$^a$ and Igor Pesando$^a$ }

\end{center}

\vspace*{4.0ex}
\centerline{ \it \small $^a$ Dipartimento di Fisica, Universit\`{a} di Torino, 
  and I.N.F.N., Sezione di Torino}
  
\centerline{ \it \small 
  Via P.\ Giuria 1, I-10125 Torino}
  
\vspace*{1.0ex}
\centerline{\small E-mail: dripto.biswas@to.infn.it,igor.pesando@to.infn.it}

\vspace*{5.0ex}

\centerline{\bf Abstract} \bigskip
We show that on shell DDF amplitudes are on shell \lc amplitudes and
that Mandelstam maps emerge naturally with a precise normalization and
are intrinsic to the DDF states.

Off shell DDF and Mandelstam amplitudes \`a la Kaku-Kikkawa  differ.


%
Underway we give a very explicit formula for the conformal
transformation of a generic vertex in the form of a compact generating
function for free theories.

\vfill

\vfill \eject
\baselineskip18pt

\tableofcontents

\section{Introduction and summary}

A hallmark of string theory is the presence of an infinite number of
higher-mass spin excitations in its spectrum, which fill the
higher-dimensional representations of the Poincaré group. String
theory offers a consistent description of these massive
modes.
However, while the interactions of massless modes are well understood,
the study of amplitudes involving higher-mass string modes remains
relatively underexplored (see
however \cite{Mitchell:1990cu,Manes:2001cs,Manes:2003mw,Manes:2004nd}
for a discussion of string form factors,
\cite{Iengo:2002tf,Iengo:2003ct,Chialva:2003hg,Chialva:2004ki,Chialva:2004xm,Iengo:2006gm,Iengo:2006if,Iengo:2006na} for the discussion on the stability of massive states which is relevant for the correspondence principle and
\cite{Arduino:2020axy} for an explanation of divergences in temporal string orbifolds due to massive states).

In the 70s, the Del Giudice, Di Vecchia, and Fubini operators, commonly named  DDF operators, were introduced to describe excited massive string states in bosonic string theory \cite{DelGiudice:1971yjh,Ademollo:1974kz}.
An important feature of these operators is that they commute with the generators of the Virasoro algebra. 
This property allows for the generation of the complete Hilbert space
of non null (non BRST exact) physical states by applying an arbitrary
combination of them to the ground state.
Generic three point amplitudes involving such states of arbitrary
string level were computed \cite{Ademollo:1974kz} and subsequently,
the framework was extended to the fermionic Neveu-Schwarz
model \cite{Hornfeck:1987wt}.
DDF operators play a crucial role in constructing string coherent
states, which are proposed as the Conformal Field Theory (CFT)
description of macroscopic objects, potentially identified with cosmic
strings \cite{Hindmarsh:2010if,Skliros:2011si}.
Recently, tree level scattering amplitudes involving DDF- operators of
exited strings have been
explored \cite{Bianchi:2019ywd,Rosenhaus:2021xhm,Firrotta:2022cku,Hashimoto:2022bll,Das:2023xge,Savic:2024ock,Firrotta:2024qel,Firrotta:2024fvi,Bhattacharya:2024szw,Biswas:2024mdu}.

In particular in \cite{Biswas:2024mdu} we have constructed the
generating functional that provides the generating function of
correlation functions for an
arbitrary number $N$ of DDF vertices, the $N$ Reggeon in the old
parlance
\cite{DiVecchia:1986jv,DiVecchia:1986uu,DiVecchia:1986mb,DiVecchia:1987ew}.
This result has been obtained
by constructing a Sciuto-Della Selva-Saito like (SDS)
vertex \cite{Sciuto:1969vz,DellaSelva:1970bj} to describe the
interaction of DDF coherent states, as introduced in
\cite{Hindmarsh:2010if,Skliros:2011si} in the context of cosmic strings.

In the quoted paper we derived this generating function both in the
standard formulation of the DDF operators and in their framed version,
as recently proposed in \cite{Biswas:2024unn}.
Generic DDF correlation functions are subsequently derived by taking
derivatives
\cite{Hindmarsh:2010if,Pesando:2012cx,Pesando:2014owa} with respect to the polarizations of the SDS vertex.

In this paper we rewrite the result obtained in \cite{Biswas:2024mdu}
in a more conventional way, i.e. using auxiliary operators acting in auxiliary
Fock spaces.
While this may seem simply a question of taste it paves the way to a
better understanding of the meaning of the DDF correlators.
We show in facts that on shell open string tree level DDF correlators
are exactly the same of on shell open string tree level \lc
correlators computed by
Mandelstam \cite{Mandelstam:1973jk,Mandelstam:1985ww,Mandelstam:1986nb}.
Explicitly this relation is given in eq. \eqref{eq:main_formula}
(and eq. \eqref{eq:derivation_DDF_lc_Reggeon_tree_level} for the Reggeon)
which we rewrite here
\begin{align}
  \Big \langle
  &
  V_{cov}(x_1;\, |DDF_1\rangle )\,
  V_{cov}(x_2;\, |DDF_2\rangle )
  \dots
  V_{cov}(x_N;\, |DDF_N\rangle )
  \Big \rangle_{UHP}
  \,
  \langle c(x_1)\, c(x_2)\, c(x_N) \rangle 
  \nonumber\\
  =&
  \cN(\{\xr\})\,
  \Big \langle
  m_{\sN 1} \circ V_{\lc}(0;\, \phi(|DDF_1\rangle) )\,
  \dots
  m_{\sN N} \circ V_{\lc}(0;\, \phi(|DDF_N\rangle) )\,
  \Big \rangle_{UHP}
  \nonumber\\
  &
  \times
  \delta(\sum_\RR k_{\sN \RR -})\,
  \delta(\sum_\RR k_{\sN \RR +})
  ,
  \label{eq:main_formula0}
\end{align}
where $\phi(|DDF\rangle)$ is the state in the \lc formalism  which
corresponds in a natural way to the DDF state $|DDF\rangle$ and it is
expressed using the \lc transverse string coordinates.

\begin{figure}[!hbt]
  \centering
  \includegraphics[width=0.9\textwidth]{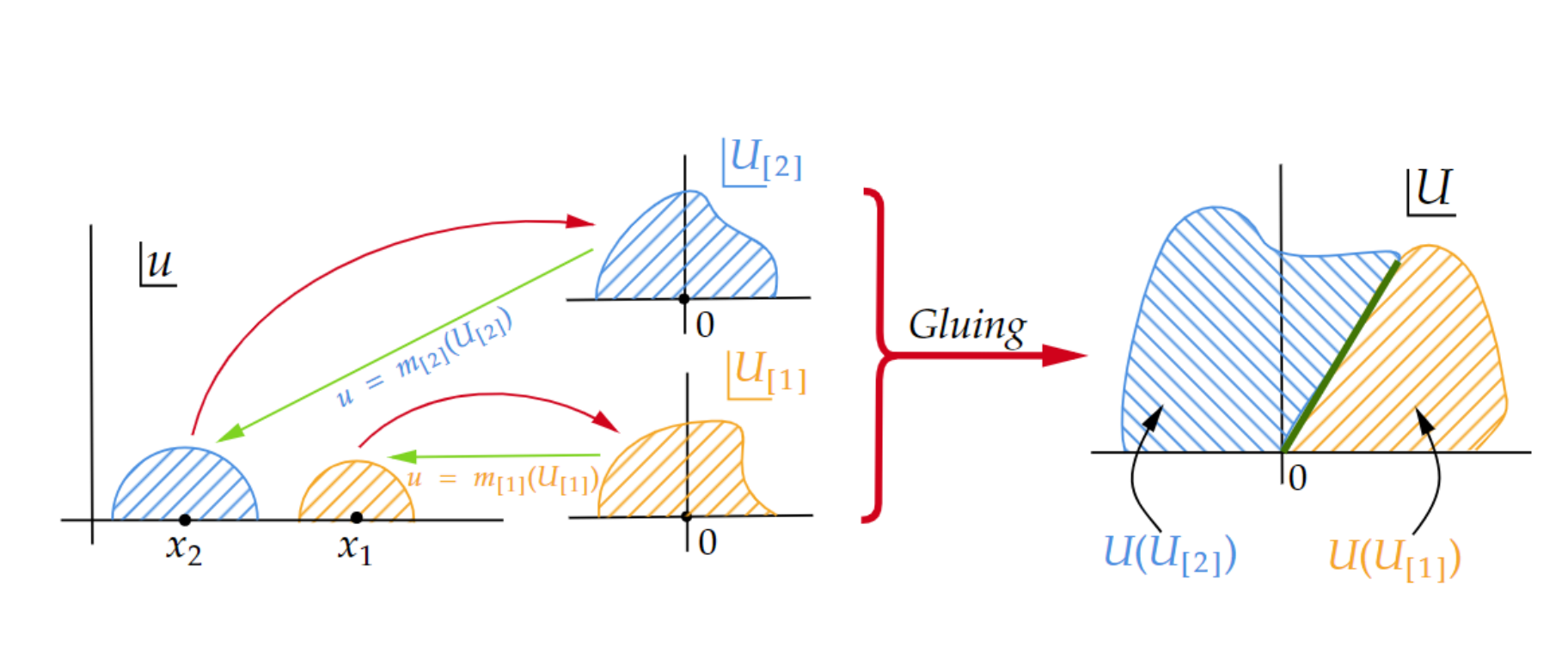}
  \caption{
In this picture we show the upper half plane with coordinate $u$
  on which we compute the DDF correlators. 
We show also how $U_{\sN \RR}$ and $m_{\sN \RR}$ arise and
  how they are connected with
  the upper half plane with cuts described by  coordinate $U$
  (when the two shown states have $\uk_{\sN 1}^+, \uk_{\sN 2}^+>0$ so
  that $U_{\sN \RR}=0$ is mapped to $U=0$ otherwise it would be mapped
  to $U=\infty$).
    }
  \label{fig:Ur_mr_globalU}
\end{figure}

In the previous formula $u=m_{\sN \RR}(U_{\sN \RR})$ ($\RR=1, \dots N$)
are the Mandelstam
maps (see figure \ref{fig:Ur_mr_globalU})
which are the local inverse of 
\begin{equation}
  \Ulc \RR(u)
  =
  (u -\xr)\,
  \prod_{\SS=1}^{\RR-1} (\xs -u)^{\frac{\uk_{\sN \SS}^+}{\uk_{\sN \SS}^+}}\,
  \prod_{\SS=\RR+1}^{N} (u- \xs)^{\frac{\uk_{\sN \SS}^+}{\uk_{\sN \SS}^+}}\,
  ,
  \label{eq:equation_for_local_patch_Mandelstam_surface00}
\end{equation}
with $\uk_{\sN \RR}^+$ the framed DDF momentum which corresponds to
$k\cdot p_{\sN \RR}$ in the usual notation ($k$ being the null
momentum entering the DDF operator and
$p_{\sN \RR}$ the tachyonic momentum of the $r$th state). 

These maps are used by Mandelstam to map charts (which corresponds to
strings and have coordinates $  \Ulc \RR$)
of the upper half plane with cuts (which has coordinate $U$)
to the upper half plane (which has coordinate $u$).
The global mapping between the upper half plane with cuts
and the upper half plane is given in eq. \eqref{eq:equation_for_Mandelstam_surface0}.
Actually he uses the strip with cuts with coordinate $\rho$.
The strip has width $\alpha$
There is however a one to one correspondence
between
the strip with cuts and 
the upper half plane with cuts
given by $U=e^{\rho \pi/ \alpha}$.
The upper half plane with cuts or the strip
describes the \lc open string interactions.

Using DDF amplitudes these maps arise in a natural way even if they are hidden.
These results essentially but not exactly agree with what obtained
in \cite{Erler:2020beb,Baba:2009kr,Baba:2009ns}
in the context of string field theory.
With a scaling difference in the Mandelstam maps.
The maps used in \cite{Erler:2020beb,Baba:2009kr,Baba:2009ns}
are such that the interactions are at $U_{\sN \RR}=1$, i.e. the unit
half disk which describes locally an ingoing or outgoing string
is without any singularity in the interior of the half disk.
This is not true with the maps which emerge from the direct
computation of the DDF amplitudes.
However these maps arise also from using Witten string field theory
maps on DDF states and therefore the existence of singularities in the
unit half disk is intrinsic to DDF or \lc states.
We notice finally that the maps obtained here in a natural way
do not give raise to problematic ``negative stubs''
noted in \cite{Erler:2020beb}.
Conceptually there is a difference in the approaches.
Here we start from off shell DDF states as discussed in \cite{Biswas:2024unn}
and we use them as (a subset of) a basis of string states
and then we  compute the off shell amplitude for $N=3$ states
according to string field theory recipes.
From this amplitude we read the Mandelstam maps.
Differently in \cite{Baba:2009kr,Baba:2009ns} they assume a given
expression for the Mandelstam maps.

While the mapping between on shell DDF amplitudes and on shell \lc
amplitudes works perfectly it seems that there are some issues in the
mapping between off shell DDF amplitudes as given by some string field
theory formulation and Kaku-Kikkawa \lc string field theory.
It seems however that these issues may be cured if a slightly
different vertex is used.
This is however strange since the vertex used in Kaku-Kikkawa \lc
string field theory is the vertex used by Mandelstam to check the
$N=4$ tachyons amplitude factorization. 

Another result is a very explicit formula for free theories.
It gives the effect of a conformal transformation $u=f(U)$ on a
generic vertex (both primary and not primary) of a free theory.
It is given in the form of a generating function\footnote{
For example
$: (\partial^3L)^2\, \partial L\, e^{i k L}:(x)
\Rightarrow
\left.
\frac{\partial^3}{\partial \lambda_3^2\, \partial \lambda_1}
\right|_{\lambda_0=k,\, \lambda_{n>0}=}
$.
}and follows from the
SDS conformal transformation.
It is given by eq. \eqref{eq:f_circ_SDS} for the operatorial
formulation or
\begin{align}
  f
  \circ
  &
  :\,
  e^{
    i \shap
    \left[
      \lambda_0  L(X)
    + 
    \sum_{n=1}^\infty \frac{\lambda_n}{(n-1)!}\, \partial_X^n  L(X)
    \right]
    }\, :\,\,
=
\nonumber\\
=&
  :\,
  e^{
    i \shap
    \left[
      \lambda_0  L( f(X) )
    + 
    \sum_{n=1}^\infty \frac{\lambda_n}{(n-1)!}\, \partial_X^n  L( f(X) )
    \right]
    }\, :
\nonumber\\
\times
&
( f'(X) )^{\oh \lambda_0^2}\,
e^{
  -\frac{2}{\ap}
  \lambda_0
   \sum_{n=1}^\infty \frac{\lambda_n}{(n-1)!}\, \partial_X^n \Delta(X,Y)|_{Y=0}
}\,
e^{
  -\frac{1}{\ap}
   \sum_{n=1}^\infty \frac{\lambda_n}{(n-1)!}\,
   \sum_{m=1}^\infty \frac{\lambda_m}{(m-1)!}\,
  \partial_X^n \partial_Y^m\Delta(X,Y)|_{Y=0}
}
,
\label{eq:f_circ_SDS_coh_form0}
\end{align}
for the coherent state formulation where $L(X)$ the chiral left moving
part of the string coordinate
and
\begin{align}
\Delta(X,Y)
=&
-\hap\,
\ln \frac{f(X) -f(Y)}{ X-Y }
.
\end{align}

In the future
it would be interesting to check that this correspondence
between DDF amplitudes and \lc ones  
works at the loop level
and to explore the maps when each DDF state has a different \lc.
The main reason is that in \lc there seems to be a simple description
of the moduli space of the Riemann surfaces while in covariant
formalism it is known only locally at the level of the measure but
the global structure is not known.

The paper is organized as follows.

In section \ref{sec:Mandelstam_maps_intro} we give the general recipe
of the mapping between DDF amplitudes and \lc ones
and we give also a simple example on how it works in
section \ref{sec:example_mapping_DDF_LC}.

We then proceed step by step to uncover the Mandelstam maps.

In order to do so we review the classical but by now not so used
formalism  of the operatorial Reggeon in section \ref{sec:old_Reggeon}.
Then we rewrite the DDF Reggeon seen as a coherent state as an
operatorial Reggeon in section \ref{sec:rewrite_Reggeon}.
This allows in section \ref{sec:rewrite_Reggeon_to_show_Madelstam} to
reassemble the operatorial Reggeon in a way which lends itself to
uncover the Mandelstam maps.
In section \ref{sec:Mandelstam_for_interacting_strings} we uncover
the Mandelstam maps for the part of the Reggeon which describes the
interaction between two different stings.
The Reggeon has also a part which describes a string self interaction.
In section \ref{sec:Mandelstam_for_self_interacting_strings} we
show that these terms arise because of the conformal transformation
performed by Mandelstam maps on \lc vertices.
In particular \lc vertices are in a CFT with non vanishing central
charge and generically non primary
and the corresponding SDS vertices have non trivial conformal
transformations.
We give a very explicit formula for free theories for the conformal
transformation of a generic vertex in the form of a compact generating
function as follows from the SDS conformal transformation.
This is derived in appendix \ref{app:conf_transf_SDS_details}.

As a final step in section \ref{sec:Nomalization_factor_vs_Mandelstam}
we compare the normalization factors on shell and discuss how they
match.
Details are given in appendix \ref{app:Reggeon_N=3_comparison}.

Finally in section \ref{sec:Mandelstam_off_shell} we discuss the
differences between the DDF results and the
Mandelstam-Kaku-Kikkawa-Cremmer-Gervais approach
\cite{Kaku:1974xu,Cremmer:1974ej,Cremmer:1974jq}.



\section{Mandelstam maps from DDF amplitudes}
\label{sec:Mandelstam_maps_intro}

In this section we would like to give a more precise overview of how
on shell amplitudes with DDF states
are \lc amplitudes computed Mandelstam maps.

In particular given a covariant vertex associated to a DDF state
we can associate a \lc vertex as
\begin{equation}
  V_{cov}(x;\, |DDF\rangle )
  \rightarrow
  V_{\lc}(x;\, \phi( |DDF\rangle ) )\,
  e^{i k_{-} x_{\lc 0}^-}
  ,
  \label{eq:DDF_to_lc_vertex}
\end{equation}
where $\phi( |DDF\rangle )$ is the natural \lc state associated to
$|DDF\rangle$.
The zero modes have to be treated carefully since on the \lc when the
gauge fixing $X^+(\sigma, \tau)= \tau$ is chosen
there are only only $x_{\lc\, 0}^i$ and $p_{\lc\, 0 i}$,
$x_{\lc\, 0}^-$ and $p_{\lc\, 0 -}$
but there is no
independent $H_\lc =P_+$ zero mode 
since  $p_{\lc\, 0 +}$ is the Hamiltonian.
In particular $\phi( |DDF\rangle )$ depends on transverse momenta
$k_i$ only.

An important point is that 
$  V_{cov}(x;\, |DDF\rangle )$ is a primary conformal operator 
of dimension $h= \ap \uk_T^2$ (where $\uk_T$ is the momentum of the
associated tachyonic state)
of a CFT of total charge $c=0$
so under a conformal transformation it
transforms as
\begin{equation}
  f \circ  V_{cov}(x;\, |DDF\rangle  )
  =
  \left( \frac{d f}{d x} \right)^h\,
    V_{cov}(f(x);\, |DDF\rangle  )
,
\end{equation}
while
$  V_{\lc}(x;\, \phi( |DDF\rangle ) ) $
is generically not a primary conformal operator
of a CFT of total charge $c=D-2$
with $h_\lc$ dependent of the state $\phi( |DDF\rangle )$ and
generically very different from $h$
so its conformal transformation
\begin{equation}
  f \circ  V_{\lc}(x;\, \phi( |DDF\rangle ) )
  =
  \left( \frac{d f}{d x} \right)^{h_{\lc}}\,
  V_{\lc}(f(x);\, \phi( |DDF\rangle ) )
  +
  \dots
,
\end{equation}
involves some ``anomalous'' terms hidden in $\dots$.

The correspondence is then\footnote{
This result is very similar to the one obtained in
\cite{Erler:2020beb} but the local maps differ by a scaling factor.
In our case there is no ambiguity in getting the local maps.
}
\footnote{
Here and in the following we use
$\uk_{\sN \RR}$ for the framed DDF momenta
and
$k_{\sN \RR}$ for the \lc theory momenta.
}
\begin{align}
  \Big \langle
  &
  V_{cov}(x_1;\, |DDF_1\rangle )\,
  V_{cov}(x_2;\, |DDF_2\rangle )
  \dots
  V_{cov}(x_N;\, |DDF_N\rangle )
  \Big \rangle_{UHP}
  \,
  \langle c(x_1)\, c(x_2)\, c(x_N) \rangle 
  \nonumber\\
  =&
  \cN(\{\xr\})\,
  \Big \langle
  m_{\sN 1} \circ V_{\lc}(0;\, \phi(|DDF_1\rangle) )\,
  \dots
  m_{\sN N} \circ V_{\lc}(0;\, \phi(|DDF_N\rangle) )\,
  \Big \rangle_{UHP}
  \nonumber\\
  &
  \times
  \delta(\sum_\RR k_{\sN \RR -})\,
  \delta(\sum_\RR k_{\sN \RR +})
  .
  \label{eq:main_formula}
\end{align}
In the previous expression
$  \cN(\{\xr\})$ is the factor that in path integral approach
and {\sl critical dimension} comes
from the functional determinant of the $2d$ Laplacian and the Jacobian
which arises from the change of coordinates from $\tau_E + i \sigma$
to upper plane ones (\cite{Mandelstam:1973jk}
and \cite{Green:1987mn} vol. 2
and \cite{Baba:2009ns} for a more recent approach).

We have also introduced the two
$  \delta(\sum_\RR k_{\sN \RR -})\,\delta(\sum_\RR k_{\sN \RR +})$.
While from the covariant point of view this seems completely arbitrary
and strange this is not so from the \lc point of view.
In facts
$\delta(\sum_\RR k_{\sN \RR -})= \delta(\sum_\RR k_{\sN  \RR}^+)$
comes from the factor $e^{i k_{\sN \RR -} x_{\lc\, 0}^-}$ in eq. \eqref{eq:DDF_to_lc_vertex}.
And $\delta(\sum_\RR k_{\sN \RR +})= \delta(\sum_\RR k_{\sN  \RR}^-)$
comes from computing the path integral over the
infinite \lc time interval since the \lc Hamiltonian is $H_\lc =P_+$.
One could wonder whether this $\delta$ along with the other factors
come from the highly non trivial natural piece $e^{i k_{\sN \RR -}
  \hat X_{\lc\, \nzm}^-}$
where $\hat X_{\lc\, \nzm}^-$ is a composite operator.

\begin{figure}[!hbt]
  \centering
  \def\svgwidth{200pt}
\begingroup%
  \makeatletter%
  \providecommand\color[2][]{%
    \errmessage{(Inkscape) Color is used for the text in Inkscape, but the package 'color.sty' is not loaded}%
    \renewcommand\color[2][]{}%
  }%
  \providecommand\transparent[1]{%
    \errmessage{(Inkscape) Transparency is used (non-zero) for the text in Inkscape, but the package 'transparent.sty' is not loaded}%
    \renewcommand\transparent[1]{}%
  }%
  \providecommand\rotatebox[2]{#2}%
  \newcommand*\fsize{\dimexpr\f@size pt\relax}%
  \newcommand*\lineheight[1]{\fontsize{\fsize}{#1\fsize}\selectfont}%
  \ifx\svgwidth\undefined%
    \setlength{\unitlength}{355.28701823bp}%
    \ifx\svgscale\undefined%
      \relax%
    \else%
      \setlength{\unitlength}{\unitlength * \real{\svgscale}}%
    \fi%
  \else%
    \setlength{\unitlength}{\svgwidth}%
  \fi%
  \global\let\svgwidth\undefined%
  \global\let\svgscale\undefined%
  \makeatother%
  \begin{picture}(1,0.65787656)%
    \lineheight{1}%
    \setlength\tabcolsep{0pt}%
    \put(0,0){\includegraphics[width=\unitlength,page=1]{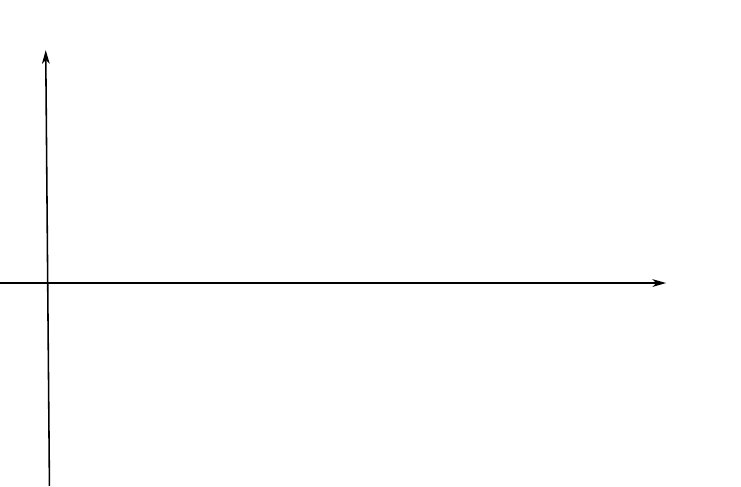}}%
    \put(0,0.63434627){\makebox(0,0)[lt]{\lineheight{1.25}\smash{\begin{tabular}[t]{l}$arg \hat U$\end{tabular}}}}%
    \put(0.9499362,0.27548147){\makebox(0,0)[lt]{\lineheight{1.25}\smash{\begin{tabular}[t]{l}$x$\end{tabular}}}}%
    \put(0,0){\includegraphics[width=\unitlength,page=2]{arg_U.pdf}}%
    \put(0.73883919,0.31770084){\makebox(0,0)[lt]{\lineheight{1.25}\smash{\begin{tabular}[t]{l}$x_1$\end{tabular}}}}%
    \put(0.14776784,0.31770084){\makebox(0,0)[lt]{\lineheight{1.25}\smash{\begin{tabular}[t]{l}$x_6$\end{tabular}}}}%
    \put(0.46441327,0.40213963){\makebox(0,0)[lt]{\lineheight{1.25}\smash{\begin{tabular}[t]{l}$x_5$\end{tabular}}}}%
    \put(0.59107143,0.40213963){\makebox(0,0)[lt]{\lineheight{1.25}\smash{\begin{tabular}[t]{l}$x_4$\end{tabular}}}}%
    \put(0.56996172,0.04327492){\makebox(0,0)[lt]{\lineheight{1.25}\smash{\begin{tabular}[t]{l}$x_3$\end{tabular}}}}%
  \end{picture}%
\endgroup%

  \caption{An example of the  argument of
    $\hat U=e^{i\phi_m} U^{\pi/(\phi_M-\phi_m)}$.}
  \label{fig:raw_U_phase}
\end{figure}

Finally we have introduced the fundamental
functions $u=m_{\sN \RR}(\Ulc \RR)$ \footnote{
We use the notation $m_{\sN \RR}$ even if it could be confused with a
mass in order to stress that it is the Mandelstam map and that is the
reason of the $m$ and it takes
value in the global coordinate $u$ which is a lower letter. 
}which are the Mandelstam maps which map local
coordinates $\Ulc \RR$ of the \lc $\RR$th vertex to the global upper
plane coordinate $u$.
The local coordinates $\Ulc \RR$ are associated to an open covering of
the Riemann surface with boundary $\Sigma_N( \{x_\RR, \uk^+_{\sN \RR} \})$
which is defined by the global equation
\begin{equation}
  \left[ \glU(u) \right]^{\frac{\phi_M-\phi_m}{\pi}}
  =
  e^{-i \phi_m}\,
  \prod_{s=1}^{N}
  (u -\xs )^{\sdap \uk^+_{\sN \SS} }
  =
  e^{-i \phi_m}\,
  \hat U
  ,
  \label{eq:equation_for_Mandelstam_surface0}
\end{equation}
where both $u$ and $U$ take values in the upper half plane while
$\hat U$ is a partial multiple cover of the upper half plane.
In the previous equation we have defined
\begin{align}
  \phi(x)
  =&
  \pi \sum_{\SS=1}^N\,
  \theta(\xs -x)\, \sdap \uk^+_{\sN \SS}
  ,
  \nonumber\\
  \phi_m = \min \phi(x),
  ~~&~~
  \phi_M = \max \phi(x),
  ,
\end{align}
so that $0 \le arg(U) \le \pi$.
See figure \ref{fig:raw_U_phase} for an example which explains the
normalizations used.

Explicitly the local coordinates $\Ulc \RR$ are
\begin{equation}
  \Ulc \RR(u)
  =
  \left[
    e^{i \pi \sum_{\SS=1}^{\RR-1} \sdap \uk_{\sN \RR}^+} \glU(u)
    \right]^{ \frac{1}{\sdap \uk_{\sN \RR}^+} }
  =
  (u -\xr)\,
  \prod_{\SS=1}^{\RR-1} (\xs -u)^{\rho_{\SS \RR}}\,
  \prod_{\SS=\RR+1}^{N} (u- \xs)^{\rho_{\SS \RR}}\,
  ,
  \label{eq:equation_for_local_patch_Mandelstam_surface0}
\end{equation}
where $\uk_{\sN \RR}^+ = \uk_{T \sN \RR}^+$ are the $+$ component of
the $\RR$th vertex momentum
and
\begin{equation}
  \rho_{\SS \RR}
  =
  \frac{ \uk_{T \sN \SS}^+ }{ \uk_{T \sN \RR}^+ }
  =
  \frac{ \uk_{ \sN \SS}^+ }{ \uk_{ \sN \RR}^+ }
  .
\end{equation}
The definition is such that an open half disk around $u=\xr$
$|u-\xr|< min_{s\ne r}\{|\xs-\xr|\}$ is mapped to the upper half plane
of $  \Ulc \RR$.
In particular singular points may occur for $| \Ulc \RR |<1$.
For example for $N=3$ the interaction point $u_I$ defined by
$\left. \frac{d\Ulc \RR}{d u} \right|_{u=u_I}=0$
is such that
$  |\Ulc \RR(u_I)|^{\sdap \ukp_{\sN \RR}} = 
\left| \frac{\ukp_{\sN 1}}{\ukp_{\sN 1}+\ukp_{\sN 2}} \right|^{\sdap \ukp_{\sN 1}}
\left| \frac{\ukp_{\sN 2}}{\ukp_{\sN 1}+\ukp_{\sN 2}} \right|^{\sdap \ukp_{\sN 2}}
$.

For later use we record that
\begin{align}
  \frac{d \Ulc{ \RR}(z)}{d u}  \Biggr|_{u=\xr}
  =&
  \prod_{\SS=1}^{\RR-1} (x_{\SS \RR})^{\rho_{\SS \RR}}
  \prod_{\SS=\RR+1}^{N} (x_{\RR \SS})^{\rho_{\SS \RR}}
  .
  \label{eq:Uhat}
\end{align}

The function $u=m_{\sN \RR}(\Ulc \RR)$ is the local inverse
of eq. \eqref{eq:equation_for_local_patch_Mandelstam_surface0} so that
$\xr=m_{\sN \RR}(\Ulc \RR=0)$.


\subsection{An explicit example of the correspondence}
\label{sec:example_mapping_DDF_LC}
Before describing how the previous correspondence wee would like to
give the simplest example of the same.

Let us consider the tachyons.
In this case we have
\begin{align}
  V_{cov}(x; |k_{T \mu}\rangle )
  =&
  :\, e^{i 2 k_{T \mu}\, \uL^\mu(x)}\, :
  ~~
  \rightarrow
  ~~
  V_{\lc}(X; |k_{T i}\rangle )
  =
  :\, e^{i 2k_{T i}\, \uLlc^i(x)}\, :
  ,
\end{align}
where $X$ is the \lc local coordinate which differs from the DDF local coordinate.
It is worth noticing that the fields used in \lc are only the
transverse ones and therefore the CFT has non trivial central charge.

In the previous case it happens that $  V_{\lc}(X; |k_{T i}\rangle )$
is a primary operator with conformal dimension
$h_\lc = \ap k_{T i} k_{T}^{ i} $ but this is an exception.
It then follows that
\begin{align}
  m_{\sN \RR} \circ V_{\lc}( 0;\, | k_{T\sN \RR}\rangle) 
  =&
  \left.  \left( \frac{d m_{\sN \RR}}{ d U_{\sN \RR}}
  \right)^{\ap k_{T \sN \RR i} k_{T \sN \RR}^{ i} }  \right|_{U_{\sN \RR}=0}\,
  V_{\lc}( \xr=m_{\sN \RR}(0) ;\, | k_{\sN \RR} \rangle)
  .
\end{align}

Then the correspondence reads
\begin{align}
  \prod_{\RR < \SS } &
  (\xr - \xs)^{\dap k_{T \sN \RR \mu} k_{T \sN   \SS}^{ \mu}}\,
  (x_1-x_2) (x_1-x_N) (x_2-x_N)
  \delta^{D-2}(\sum_\RR k_{T \sN \RR i})\,
  \delta(\sum_\RR k_{T \sN \RR -})\,
  \delta(\sum_\RR k_{T \sN \RR +})
  \nonumber\\
  =&
  \cM(\{\xr\})\,
  \times
  \prod_{\RR}
  \left.  \left( \frac{d m_{\sN \RR}}{ d U_{\sN \RR}}
  \right)^{\ap k_{T \sN \RR i} k_{T \sN \RR}^{ i} }  \right|_{U_{\sN \RR}=0}\,
  \prod_{\RR < \SS } (\xr - \xs)^{\dap k_{T \sN \RR i} k_{T \sN
      \SS}^{ i}}\,
  \delta^{D-2}(\sum_\RR k_{T \sN \RR i})\,
  \nonumber\\
  &
  \times
  \delta(\sum_\RR k_{T \sN \RR -})\,
  \delta(\sum_\RR k_{T \sN \RR +})
  .
\end{align}

Now we can use eq. \eqref{eq:Uhat} and
evaluate 
\begin{align}
  \prod_\RR
  &
  \left.
  \left( \frac{d m_{\sN \RR}}{ d U_{\sN \RR}}
  \right)^{\ap k_{T \sN \RR i} k_{T \sN \RR}^{ i} }
  \right|_{U_{\sN     \RR}=0}
  = 
  \prod_\RR
  \left.
  \left( \frac{ d U_{\sN \RR}}{d u} \right)
  ^{-\ap (k_{T \sN \RR i} k_{T   \sN \RR}^{ i} +m_{T \sN \RR}^2 )}
  \right|_{u=\xr}
  \times
  \prod_\RR
  \left.
  \left( \frac{ d U_{\sN \RR}}{d u} \right)^{+\ap m_{T \sN \RR}^2 }
  \right|_{u=\xr}
  \nonumber\\
  =&
  \prod_{\RR=1}^N
  \left(
  \prod_{\SS=1}^{\RR-1} (\xs -\xr)^{ \frac{ \uk_{T \sN \SS}^+ }{ \uk_{T \sN \RR}^+ }}
  \prod_{\SS=\RR+1}^{N} (\xr-\xs)^{ \frac{ \uk_{T \sN \SS}^+ }{ \uk_{T \sN \RR}^+ }}
  \right)^{-\ap (k_{T \sN \RR i} k_{T   \sN \RR}^{ i}+m_{T \sN \RR}^2) }
  \times
  \prod_\RR
  \left.
  \left( \frac{ d U_{\sN \RR}}{d u} \right)^{+\ap m_{T \sN \RR}^2 }
  \right|_{u=\xr}
  \nonumber\\
  =&
  \prod_{\RR=1}^N
  \left(
  \prod_{\SS=1}^{\RR-1} (\xs -\xr)^{- \dap { \uk_{T \sN \SS}^+ }{ \uk_{T \sN \RR}^- }}
  \prod_{\SS=\RR+1}^{N} (\xr-\xs)^{- \dap {\uk_{T \sN \SS}^+ }{ \uk_{T \sN \RR}^- }}
  \right)
  \times
  \prod_\RR
  \left.
  \left( \frac{ d U_{\sN \RR}}{d u} \right)^{+\ap m_{T \sN \RR}^2 }
  \right|_{u=\xr}
    ,
\end{align}
where we used $k^- = \frac{k_i k^i + m^2}{2 k^+}$.
The first factor in the last line completes the \lc exponents
to
$\dap k_{T \sN \RR \mu} k_{T \sN   \SS}^{ \mu}
=
-\dap k_{T \sN \RR +} k_{T \sN   \SS -}
-\dap k_{T \sN \RR -} k_{T \sN   \SS +}
+\dap k_{T \sN \RR i} k_{T \sN   \SS i}
$
while the second one (when the amplitude is on shell,
i.e. with $+\ap m_{T \sN \RR}^2=-1$)
along with $\cM(\{\xr\})\,$ makes the ghost contribution.

\section{Reminder on the usual operatorial formulation of the
  Reggeon vertex}
\label{sec:old_Reggeon}

Let us start remembering that the $N$-reggeon vertex can be obtained
from the SDS vertex.

Given any state, physical or unphysical, $|\phi\rangle$ in an Hilbert
space $\cH$
we can explicitly compute the
corresponding operator in the different Hilbert space $\widetilde\cH$.
This vertex operator is
$ V[\tilde X](x;\, |\phi\rangle) $
and can be easily computed 
using the Sciuto-Della Selva-Saito vertex as
\begin{equation}
  \widetilde \cS(x) |\phi\rangle
  =
  V[\tilde X](x;\, |\phi\rangle)
  .
\end{equation}
Explicitly the Sciuto-Della Selva-Saito vertex for the covariant
formulation is given by
\begin{align}
  \widetilde \cS(x)
  &=
  \langle x^\mu=0, 0_{a^\mu} |
  :
  e^{
    \frac{i}{\sdap}
    g_{\mu \nu}
    \sum_{n=0}^\infty \frac{\alpha^\mu_{n}}{n!}
    \partial^n_x \tilde X^\nu(x, \bar x)
  }
  :
  \nonumber\\
  &=
  \langle x^\mu=0, 0_{ a^\mu} |
  :
  e^{
    -\frac{2}{\ap}
    g_{\mu \nu}
    \oint_{z=0,\, |z|<|x|} \frac{d z}{ 2\pi i}
    \partial_z L^\mu(z)
     \tilde L^\nu(x+z)
  }
  :
  \nonumber\\
  =&
  \langle x^\mu=0, 0_{a^\mu} |
  :
  e^{
    i \ishap
    g_{\mu \nu}
    \left(
   \alpha^\mu_0 \tilde L^\nu(x)
    +
    \sum_{n=1}^\infty \frac{\alpha^\mu_{n}}{n!}
    \partial^n_x \tilde L^\nu(x)
    \right)
  }
  :
~~~~\mbox{when }x>0
  ,
  \label{eq:SDS1}
\end{align}
where the normal ordering is acting \wrt $\tilde \alpha$ operators
since only for them there are creators and annihilators,
 the vacuum $\langle 0_a|$ is the one for the $\alpha$ operators
 acting in $\cH$ and similarly
 $\langle x=0|$ is the null eigenstate of $x_0$.

The $N$ Reggeon, i.e. the generating function for all the tree level amplitudes
is then
\begin{equation}
  \langle V_N(x_1, \dots x_N) |
  =
  \langle \tilde 0 |\,
  \widetilde \cS_{\sN 1}(x_1)
  \dots
  \widetilde \cS_{\sN N}(x_N)\,
  | \tilde 0\rangle_{\widetilde{UHP}}
  ,
  ~~~~~
  |x_1|>\dots>|x_N|
  ,
  \label{eq:derivation_Reggeon_tree_level}
\end{equation}
and it is such that any correlator can be computed as
\begin{align}
  \langle V_N(x_1, \dots x_N) |
  \left(
  |\phi_{\sN 1}\rangle \otimes \,
  \dots
  |\phi_{\sN N}\rangle
  \right)
  =&
  \Big \langle \tilde 0 \Big |\,
  \tilde V(x_1 ; \phi_{\sN 1})
  \dots
  \tilde V(x_N ; \phi_{\sN N})\,
  \Big |\tilde 0 \Big \rangle_{\widetilde{UHP}}
  ,
\end{align}
where   $\tilde V(x_\RR ; |\phi_{\sN \RR} \rangle )$ is the vertex
operator for the 
emission of the state $|\phi_{\sN \RR}\rangle$ in the Fock space
$\tilde \cH$
and the correlators are computed in the UHP.

An easy computation gives
\begin{align}
  \cV_N^{(cov)}
  &(\{ \xr \})
  =
  \langle V_N(x_1, \dots x_N)|
  =
  \nonumber\\
  &
  \left(
  \otimes_{\RR=1}^N
  \langle x^\mu_{\sN \RR 0}=0; 0_{a^\mu_{\sN \RR}} |
  \right)
  \,
  \prod_{1\le \RR < \SS \le N}
  e^{-\frac{2}{\ap}
    \oint_{z_{\sN \RR}=0}
    \oint_{z_{\sN \SS}=0}
    \frac{d z_{\sN \RR}}{2\pi i}
    \frac{d z_{\sN \SS}}{2\pi i}
    \partial L^{(+) \mu}_{\sN \RR}(z_{\sN \RR})
    \partial L^{(+) \mu}_{\sN \SS}(z_{\sN \SS})
   \ln\left( z(z_{\sN \RR}) - z(z_{\sN \SS}) \right)
  }
  \nonumber\\
  &
  \phantom{\otimes_{r=1}^N}
  \times
  \delta^{D}(\sum_{\RR=1}^N \alpha_{\sN \RR 0}^\mu)
  ,
 \label{eq:N Reggeon}
\end{align}
where we defined the map
\begin{equation}
    z(\Zlc \RR) = \Zlc \RR +x_\RR
,
\end{equation}
for which $    z(\Zlc \RR=0) =\xr$.
This can be interpreted as the map between the local doubled
coordinate $\Zlc {\RR }$ associated with the covariant vertex
and the doubled global coordinate $z$.
It is the local inverse of the map
\begin{equation}
  \Zlc \RR(z) = z - \xr
  \label{eq:usual_local_global}
.
\end{equation}
We have written doubled because the true map for the open string would
read
\begin{equation}
    u(\Ulc{\RR }) = \Ulc{ \RR } +\xr
,
\end{equation}
where both $u$ and $u_{\sN \RR }$ are defined on the UHP but we need
to double their domain of definition to the whole complex plane in order to be
able to perform the contour integrals.
In this case the Riemann surface is simply
\begin{equation}
  U=Z
  .
\end{equation}

\section{Recovering Mandelstam map 1: mapping coherent state
  parameters $\lambda$ to operator}
\label{sec:rewrite_Reggeon}

We need to compare the usual expression
for the Reggeon using operators in \eqref{eq:N Reggeon}
with the one using $\lambda_{\sN r n}^\mu$:
%
%
%
%
\begin{align}
  {\cV}_N^{(DDF)}
  & (\{ \xr \})
  =
  \Big \langle \up^i=\up^-=0 \Big|
  \nonumber\\
  %
  %
  \prod_{\TT=1}^N
  \Bigg\{
  \exp
  &
  \Bigg(
  \sum_{i=2}^{D-2}
  \sum_{n=1}^\infty
  \lambda^i_{\sN \TT n}
  \oint_{z=0} \frac{d z}{ 2\pi i}
  \left(
  \sum_{\RR \ne \TT}
  \frac{\sdap \uk^i_{T \sN \RR} }{z + x_{\TT \RR}}
  +
  \sdap \uk^i_{T \sN t}
  \frac{1}{z}
  \right)
  [ \Zlc{\TT}(z +\xt) ]^{-n}\,
  e^{-i n \frac{ \ux_0^+ }{\dap \uk^+_{T \sN t}} }
  \Bigg)
  \nonumber
\end{align}
\begin{align}
  %
  \times
  \exp
  &
  \Bigg(
  \oh
  \sum_{i=2}^{D-2}
  \sum_{n_1, n_2=1}^\infty
  \sum_{r,t=1}^{N}
  \lambda^i_{\sN r n_1}
  \lambda^i_{\sN t n_2}
  \oint_{z_1=0, |z_1| > |z_2|} \frac{d z_1}{ 2\pi i}
  \oint_{z_2=0} \frac{d z_2}{ 2\pi i}
  \frac{
    e^{
      -i
      \frac{
        n_{ 1}  \ux_0^+
      }{\dap \uk^+_{T \sN r} }
      -i
      \frac{
        n_{ 2} \ux_0^+
      }{\dap \uk^+_{T \sN t} }
    }
  }{
    (z_1 -z_2 +x_{\RR \TT} )^2
  }
  \nonumber\\
  &
  \phantom{\exp}
  [ \Zlc{\RR}(z_1 +\xr) ]^{-n_1}\,
  [ \Zlc{\TT}(z_2 +\xt) ]^{-n_2}\,
  \Bigg)
  \Bigg\}
  \nonumber
\end{align}
\begin{align}
  \times
  &
  \prod_{\TT=1}^N
  e^{i  \uk_{T \sN \TT}^i \ux^i_0 - i  \uk_{T \sN \TT}^- \ux^+_0}
  \,
  \Big | \up^i=\up^-=0 \Big \rangle
  \,
  \nonumber\\
  &
  \prod_{\RR=1}^{N-1} \prod_{\TT=\RR+1}^N
  x_{\RR \TT}^{\dap \uk_{T \sN \RR} \cdot \uk_{T \sN \TT} }
  \,
  \delta\left( \sum_{\RR}^N \uk^+_{\sN \RR} \right)
  ,
  \label{eq:coherentDDF0}
\end{align}
where we use the notation
\begin{equation}
  x_{\SS \TT}\equiv \xs -\xt,
  ~~~~
  \rho_{\SS \TT}
  \equiv
  \frac{\uk^+_{T \sN \SS} }{\uk^+_{T \sN \TT}}
  ,
\end{equation}
and have introduced the doubled version of
eq. \eqref{eq:equation_for_local_patch_Mandelstam_surface0}
and defined the analytic functions in the cut complex plane $z$
\begin{equation}
    \Zlc{\RR}(z)
  =
  (z -\xr)
  \prod_{\SS=1}^{\RR-1}
  (\xs  - z)^{\rho_{\SS \RR}}
  \prod_{\SS=\RR+1}^{N}
  (-\xs  + z)^{\rho_{\SS \RR}}
  .
\label{eq:Zr zr}  
\end{equation}
They are analytic for $|z-\xr|< min_{s\ne r}\{|\xs-\xr|\}$.
However it may well happen that
$ \frac{ d \Zlc{\RR} } { d z} =0$ at one point closer
and then we cannot invert the map anymore.

For later use notice that eq. \eqref{eq:Uhat} is in the doubled version
\begin{align}
  \frac{d \Ulc{ \RR}(z)}{d u}  \Biggr|_{u=\xr}
  =
  \frac{d \Zlc{ \RR}(z)}{d z}  \Biggr|_{z=\xr}
  =&
  \prod_{\SS=1}^{\RR-1} (x_{\SS \RR})^{\rho_{\SS \RR}}
  \prod_{\SS=\RR+1}^{N} (x_{\RR \SS})^{\rho_{\SS \RR}}
  .
    \label{eq:UhatZhat}
\end{align}

We want to map the previous equation eq. \eqref{eq:coherentDDF0}
to an equation with $\alpha$s similar to eq. \eqref{eq:N Reggeon}.
In our case $\alpha^\mu$ are actually $\uA^i$ or better $\Alc^i$.
In order to do so
we have to map $\lambda$s which are associated with DDF operators
to $\Alc$s.
It is  easy to realize that we have the mapping
\begin{equation}
  \begin{cases}
  \lambda_{\sN \RR n}^i
  ~~\leftrightarrow~~
  +\frac{1}{n}\Alc_{\sN \RR n}^i
  &
  n\ge 1
  \\
  \sdap k^i_{\sN \RR }
  ~~\leftrightarrow~~
  -\Alc^i_{\sN \RR 0}
  &
  \end{cases}
  ,
  \label{eq:lambda_to_alpha_rewriting}
\end{equation}
where the $1/n$ comes from the $\Alc$ algebra
and the \lc zero modes $\Alc^i_{\sN \RR 0}$ have an index which
spans only the transverse directions and not the whole spacetime.
In facts $k^+,~ k^-$ are treated in a different way because we are already
thinking to the correspondence with the DDF case.
Finally notice that we treat all momenta in a different way \wrt non
zero modes since
there is an added $-1$ to get better equations afterwards.
The reason is that we want to get eq. \eqref{eq:interaction_an_a0}
with the factor $  -\frac{2}{\ap}$ in order to be able to write eq.
\eqref{eq:interaction_r_ne_t}.

\section{Recovering Mandelstam map 2: rewriting the Reggeon with
operators}
\label{sec:rewrite_Reggeon_to_show_Madelstam}
We are now ready to rewrite the N point DDF Reggeon using the
operators.

As a first step we eliminate all the $e^{\dots \ux_0^+}$ exponentials.
To do so  we notice that each time we take a derivative \wrt
$\lambda_{\sN \RR  n}^i$ we bring down a
$ \left( e^{ -i \frac{\ux_0^+}{ 2\ap \uk_{\sN \RR}^+ } }\right)^n $
which increases $\uk_{ T \sN \RR}^-$ by
$\frac{n}{2\ap \uk_{\sN    \RR}^+}$.
This means that when we have taken all derivatives
we have to replace the tachyonic $\uk_{ T \sN \RR}^-$ with the state momentum
$\uk_{\sN \RR}^-$.

So we get rid of the exponential simply by dropping them and making
the substitution
$\uk_{ T \sN \RR}^- \rightarrow \uk_{\sN \RR}^-$
in the factor $ e^{i  \uk_{T \sN \RR}^i \ux^i_0 - i  \uk_{T \sN \RR}^- \ux^+_0}$.

Then we can compute the zero modes expectation value and
get a conservation
$\delta^{D-2}(\sum_\RR\, \uk_\RR^i)$.
Finally use the $\lambda \rightarrow \Alc$ mapping
to write
$\delta^{D-2}(\sum_\RR\, \Alc_{\sN \RR 0}^i)$.

As a second step
we rewrite $\uk_{ T \sN \RR}^-$ in the exponents of
$  x_{\RR \TT}^{\dap \uk_{T \sN \RR} \cdot \uk_{T \sN \TT} }$
in term of the true momenta $\uk_{\sN r}$.

Moreover we want to get rid of $\uk_{\sN r\, +}= -\uk_{\sN r}^-$ since
they are dependent on the other momenta in the Hamiltonian \lc formulation.
To this purpose we notice that
\begin{align}
  \uk_{T \sN r}^i &= \uk_{\sN r}^i,
  &
  \uk_{T \sN r}^+ &= \uk_{\sN r}^+,
  \nonumber\\
  &
  \uk_{T \sN \RR}^- =
  \frac{ \uk_{\sN \RR}^i \uk_{\sN \RR}^i +m_{T \sN \RR}^2 }{2 \uk_{\sN \RR}^+}
  ,
  &&
\end{align}
where we have introduced an ``off shell'' tachyonic mass $m_{T \sN \RR}^2$
which may vary for each particle.
Then we get
\begin{align}
  \prod_{\RR=1}^{N-1} \prod_{\TT=\RR+1}^N\,
  &
  x_{\RR \TT}^{\dap \uk_{T \sN \RR} \cdot \uk_{T \sN \TT} }
  =
  \prod_{\RR=1}^{N-1} \prod_{\TT=\RR+1}^N\,
  x_{\RR \TT}^{\Alc_{\sN \RR 0}^i  \Alc_{\sN \TT 0}^i }
  \nonumber\\
  &
  \times
  \prod_{\RR=1}^{N-1} \prod_{\TT=\RR+1}^N\,
  x_{\RR \TT}^{
    \rho_{\RR \TT}
    \left( \oh \Alc_{\sN \TT 0}^i  \Alc_{\sN \TT 0}^i + \ap m_{T \sN \TT}^2 \right)
    +
    \rho_{\RR \TT}
    \left( \oh \Alc_{\sN \RR 0}^i  \Alc_{\sN \RR 0}^i + \ap m_{T \sN \RR}^2 \right)
    }
  \nonumber\\
 = &
  \prod_{\RR=1}^{N-1} \prod_{\TT=\RR+1}^N\,
  x_{\RR \TT}^{\Alc_{\sN \RR 0}^i  \Alc_{\sN \TT 0}^i }
  \times
  \prod_{\RR=1}^{N}\,
  \Bigg[
    \left. \frac{d Z_{\sN \RR}(z)}{d z} \right|_{z=\xr}
  \Bigg]^{ -\oh \Alc_{\sN \RR 0}^i  \Alc_{\sN \RR 0}^i -\ap m_{T \sN \RR}^2 }
  .
\end{align}

The other terms are easier to rewrite.
Keeping in mind our goal of rewriting the Reggeon in term of the \lc
SDS vertexes we write 
\begin{align}
  \langle V_N^{(DDF+ghost)}&( \{ \xr \})|
  =
  \left(
  \otimes_{\RR=1}^N\,
  \langle   \ux^i_{\sN r 0}, 0_{\Alc^i_{\sN \RR}}|
  \right)
  \,
  \prod_{\RR=1}^{N}
  \Biggl\{
  \left[
    \frac{d \Zlc{ \RR}(z)}{d z}
    \Biggr|_{z=\xr}
    \right]^{
    + \ap m_{T \sN \RR}^2
  }
  \Biggr\}
  (x_1-x_2)\,(x_1-x_N)\,(x_2-x_N)
  \nonumber\\
  %
  %
  \times
  &
  \prod_{\RR=1}^{N}
  \Biggl\{
  \left[
    \frac{d \Zlc{ \RR}(z)}{d z}
    \Biggr|_{z=\xr}
    \right]^{
    -\oh \sum_i\Alc_{\sN \RR 0}^{i} \Alc_{\sN \RR 0}^{i}
    }
  \nonumber\\
  %
  %
  &
  \exp
  \Biggl[
    -
    %
    \sum_{i=2}^{D-2} \sum_{n=1}^\infty
    \frac{\Alc_{\sN \RR n}^i}{n}\,
    \Alc_{\sN \RR 0}^i\,
    \oint_{z=\xr} \frac{d z}{2\pi i}
    \frac{ \Zlc{ \RR}(z)^ {-n} }{ z -\xr} 
    \Biggr]
  \nonumber\\
  %
  &
  \exp
  \Biggl[
    \oh
    \sum_{i=2}^{D-2}\,
    \sum_{n_1, n_2=1}^\infty
    \frac{\Alc_{\sN \RR n_1}^i}{n_1}
    \frac{\Alc_{\sN \RR n_2}^i}{n_2}
    \oint_{z_1= \xr} \frac{d z_1}{2\pi i}
    \oint_{z_2= \xr} \frac{d z_2}{2\pi i}
    \frac{
      \Zlc{ r}(z_1)^ {-n_1}\, \Zlc{ r}(z_2)^ {-n_2}
    }{ (z_1 -z_2)^2 }
    \Biggr]
  \Biggr\}
  \nonumber
\end{align}
\begin{align}
  %
  %
  \times
  &
  \prod_{\RR=1}^{N-1} \prod_{\TT=\RR+1}^{N}
  x_{\RR \TT}^{
    \sum_{i=2}^{D-2} \Alc_{\sN \RR 0}^i  \Alc_{\sN \TT 0}^i 
    }
  \nonumber\\
  %
  %
  &
  \exp
  \Biggl[
    -
    \sum_{\RR, \TT=1,\, \RR\ne\TT}^N\,
    \sum_{i=2}^{D-2} \sum_{n=1}^\infty
    \frac{\Alc_{\sN \RR n}^i}{n}\,
    \Alc_{\sN \TT 0}^i\,
    \oint_{z=\xr} \frac{d z}{2\pi i}
    \frac{ \Zlc{ \RR}(z)^ {-n} }{ z -\xt} 
    \Biggr]
  \nonumber\\
  %
  %
  &
  \exp
  \Biggl\{
  \sum_{\RR=1}^{N-1} \sum_{\TT=\RR+1}^{N}\,
  \sum_{i=2}^{D-2}\,
  \sum_{n_1, n_2=1}^\infty
  \frac{\Alc_{\sN \RR n_1}^i}{n_1}
  \frac{\Alc_{\sN \TT n_2}^i}{n_2}
  \oint_{z_1= \xr} \frac{d z_1}{2\pi i}
  \oint_{z_2= \xt} \frac{d z_2}{2\pi i}
  \frac{
    \Zlc{ \RR}(z_1)^ {-n_1}\, \Zlc{\TT}(z_2)^ {-n_2}
  }{ (z_1 -z_2)^2 }
    \Biggr\}
    \nonumber
\end{align}
%
%
\begin{align}
  &
  \times
    \delta^{D-2}\left(
    \sum_{\RR=1}^N \Alc_{\sN \RR 0}^i
    \right)
    \,
    \delta\left(
    \sum_{\RR=1}^N \uk_{\sN \RR}^+
    \right)\,
    \,
    \delta\left(
    \sum_{\RR=1}^N \uk_{\sN \RR }^-
    \right)
    ,
    \label{eq:Reggeon_rewritten_v1}
\end{align}
where we have grouped the different factors having in mind the final
result.
Notice the special treatment for $\uk^\pm$.

\section{Recovering Mandelstam map 3: Mandelstam maps from
  interactions between different strings}
\label{sec:Mandelstam_for_interacting_strings}
Let us now use
the substitution rule
\begin{equation}
  \lambda_{\sN \RR n}^i
  ~~\rightarrow~~
  \frac{\Alc_{\sN \RR n}^i}{n}
  =
  -i \ishap
  \oint_{W_{\sN \RR}=0}
  \frac{d \Zlc{\RR}}{2\pi i}
  \uLlc_{\sN  \RR (\ne)}^{(+) i}(\Zlc{\RR})
  \,
  \Zlc{\RR}^{n-1}
  ,~~~~
  n>1
  \label{eq:lambda_to_alpha}
\end{equation}
where $\Zlc{\RR}$ is the local coordinate around the \lc emission vertex
and we defined as usual
\begin{equation}
  \uLlc_{\sN \RR (\ne) }^{(+) i} (\Zlc{r})
  =
  i
  \shap
  \sum_{n=1}^\infty
  \frac{\Alc_{\sN \RR n}^{i} } {n}
  \Zlc{\RR}^{-n}
  ,~~~~
  \uLlc_{\sN \RR (0) }^{i} (\Zlc{\RR})
  =
  \oh \uxlc_{\sN \RR 0}^i
  -i \ap \uplc_{\sN \RR 0}^i\, \ln(\Zlc{\RR})
  ,
\end{equation}
so that trivially
\begin{equation}
  \partial \uLlc_{\sN r (0)}^i(\Zlc{\RR})
  =
  -i \shap
  \frac{\Alc_{\sN \RR 0}^i}{ \Zlc{\RR}}
  ,~~~~
  \partial \uLlc_{\sN r (\ne)}^{(+) i}(\Zlc{\RR})
  =
  -i \shap
  \sum_{n=1}^\infty
  \frac{\Alc_{\sN \RR n}^i}{ \Zlc{\RR}^{n+1}}
.  
\end{equation}

We now perform all steps carefully for the first terms
while for the other terms we quote the result since the steps are equal.

Explicitly we exam the terms
\begin{align}
    \Alc_{ 0} \Alc_{ n}\mbox{ terms }
\Rightarrow
-\sum_{\RR=1}^{N} \sum_{\TT=1, \TT\ne \RR}^N\,
\sum_{i=2}^{D-2} 
&
  \Alc_{\sN \TT 0}^{ i}
  \oint_{z= \xr} \frac{d z}{2\pi i}\,
  \sum_{n=1}^\infty
  \frac{\Alc_{\sN \RR n}^{i}}{n}
  \Zlc{\RR}(z)^ {-n}\,
    \frac{ 1 }{ z -\xt }
    .
\end{align}
We can rewrite the expression of interest as
\begin{align}
  =
+
\frac{2}{\ap}
\sum_{\RR=1}^{N} \sum_{\TT=1, \TT\ne \RR}^N\,
\sum_{i=2}^{D-2}     
  &
  \oint_{\Zlc{\TT} =0}  \frac{ d \Zlc{\TT}} { 2\pi i}\,
  \oint_{z= \xr} \frac{d z}{2\pi i}\,
    \frac{
      \uLlc_{\sN \RR (\ne) }^{(+) i} (\Zlc{\RR}( z ) )
      \,
      \partial \uLlc_{\sN \TT (0) }^{(+) i} (\Zlc{\TT})
    }{
      z -\xt
    }
    .
\end{align}

Now we change integration variable from $z$ to $Z_{\sN r}$.
and use the fact that $z(Z_{\sN r}=0)=\xr$.
This is possible since the mapping is analytic for
$|z - \xr|< min_{s\ne r}\{|\xs -\xr|\}$ and invertible for a possibly
smaller open set determined by the zeros of
$\frac{ d \Zlc{\RR} } { d z} =0$.
Notice however that this means that we may have singularities for
$|Z_{\sN r}|<1$ and this is the major difference with
\cite{Erler:2020beb,Baba:2009kr,Baba:2009ns} and somewhat unusual for
string field theory.

We get
\begin{align}
  =
  +
  \frac{2}{\ap}
  \sum_{\RR=1}^{N} \sum_{\TT=1, \TT\ne \RR}^N\,
  \sum_{i=2}^{D-2}     
  &
  \oint_{\Zlc{\TT} =0} \frac{ d \Zlc{\TT}} { 2\pi i}\,
  \oint_{\Zlc{\RR}=0} \frac{d \Zlc{\RR}}{2\pi i}\,
    \partial \uLlc_{\sN \TT (0) }^{(+) i} (\Zlc{\TT})
    \,
    \uLlc_{\sN \RR (\ne) }^{(+) i} (\Zlc{\RR} )
      \frac{ d z }{ d  \Zlc{\RR}}
    \frac{ 1 }{
      z(\Zlc{\RR}) -\xt
    }
    \nonumber\\
    =
    +
  \frac{2}{\ap}
  \sum_{\RR=1}^{N} \sum_{\TT=1, \TT\ne \RR}^N\,
  \sum_{i=2}^{D-2}
  &
  \oint_{\Zlc{\TT} =0} \frac{ d \Zlc{\TT}} { 2\pi i}\,
  \oint_{\Zlc{\RR}=0} \frac{d \Zlc{\RR}}{2\pi i}\,
  \uLlc_{\sN \RR (\ne) }^{(+) i} (\Zlc{\RR} )
  \,
  \partial \uLlc_{\sN \TT (0)}^{(+) i} (\Zlc{\TT})
  \frac{d  }{ d  \Zlc{\RR}}
  \ln \left(
  z(\Zlc{\RR}) -\xt
  \right)
  .
\end{align}
Now we integrate by part \wrt $\Zlc{\RR}=0$.
This can be done despite the logarithm since we are integrating around
$\xr$ and $\xt\ne0$ so we do not cross the cut
and get
\begin{align}
  =
  &
  -
  \frac{2}{\ap}
  \sum_{r=1}^{N} \sum_{t=1, t\ne r}^N \sum_{i=2}^{D-2}
  \oint_{\Zlc{\TT} =0}
  \frac{ d \Zlc{\TT}} { 2\pi i}
  \oint_{\Zlc{\RR}=0} \frac{d \Zlc{\RR}}{2\pi i}
  \partial \uLlc_{\sN r (\ne) }^{(+) i} (\Zlc{\RR} )
  \,
  \partial \uLlc_{\sN t (0)}^{(+) i} (\Zlc{\TT})
  %
  \ln \left(
  z(\Zlc{\RR})  -x_t
  \right)
  \nonumber\\
  =
  &
  -
  \frac{2}{\ap}
  \sum_{r=1}^{N} \sum_{t=1, t\ne r}^N \sum_{i=2}^{D-2}
  \oint_{\Zlc{\TT} =0}  \frac{ d \Zlc{\TT}} { 2\pi i}\,
  \oint_{\Zlc{\RR}=0} \frac{d \Zlc{\RR}}{2\pi i}\,
  \partial \uLlc_{\sN r (\ne) }^{(+) i} (\Zlc{\RR} )
  \,
  \partial \uLlc_{\sN t (0) }^{(+) i} (\Zlc{\TT})
  %
  \ln \left(
  z(\Zlc{\RR})  - z(\Zlc{\TT})
  \right)
  ,
  \label{eq:interaction_an_a0}
\end{align}
where the last step is possible since
$\partial \uL_{\sN t (0) }^{(+) i} (\Zlc{\TT})$ has only a single pole
at $\Zlc{\TT}=0$ and $z(\Zlc{\TT}=0)= \xt$.

Along the same line we can consider
$\Alc_{\sN r 0}^i \Alc^i_{\sN  t 0}$ with $r \ne t$.
We get
\begin{align}
  \Alc_{\sN r 0}& \Alc_{\sN t 0}\mbox{ terms }
  \Rightarrow
  \nonumber\\
  &
  \prod_{r=1}^{N-1} \prod_{t=r+1}^N
  x_{r t}^{
    \Alc_{\sN r 0}^i  \Alc_{\sN t 0}^i 
    }
  =
  \exp\{
  \sum_{r=1}^{N-1} \sum_{t=r+1}^N
  \sum_{i=2}^{D-1}
    \Alc_{\sN r 0}^i  \Alc_{\sN t 0}^i 
    \,
    \ln( z(Z_{\sN r})  - z(Z_{\sN t}) |_{ Z_{\sN r} = Z_{\sN t} =0}
    \}
    \nonumber\\
 =&
    \exp\left\{
    -
    \frac{2}{\ap}
  \sum_{r=1}^{N-1} \sum_{t=r+1}^N
  \sum_{i=2}^{D-1}
  \oint_{Z_{\sN r}=0} \frac{d Z_{\sN r}}{2\pi i}
  \oint_{Z_{\sN t}=0} \frac{d Z_{\sN t}}{2\pi i}
  \partial \uL_{\sN r (0) }^{(+) i} (Z_{\sN r} )
  \,
  \partial \uL_{\sN t (0) }^{(+) i} (Z_{\sN t} )
  \,
      \ln \left(
      z(Z_{\sN r})  -z(Z_{\sN t})
    \right)
    \right\}
    .
      \label{eq:interaction_a0_a0}
\end{align}

Finally let us now consider
$\Alc_{\sN r n_1}^i \Alc^i_{\sN  t n_2}$ with $r \ne t$.
We get
\begin{align}
  \Alc_{\sN \RR n_1} & \Alc_{\sN \TT n_2}\mbox{ terms }
  \Rightarrow
  \nonumber\\
  +
  &
  \sum_{\RR=1}^{N-1} \sum_{t=\RR+1}^{N}
  \sum_{i=2}^{D-2} \sum_{n_1, n_2=1}^\infty
  \frac{\Alc_{\sN \RR n_1}^i}{n_1}
  \frac{\Alc_{\sN \TT n_2}^i}{n_2}
  \oint_{z_1=0} \frac{d z_1}{2\pi i}
  \oint_{z_2=0} \frac{d z_2}{2\pi i}
  \frac{
    Z_{\sN \RR}(z_1)^ {-n_1}\, Z_{\sN \TT}(z_2)^ {-n_2}
  }{ (z_1 -z_2 +x_{\RR \TT})^2 }
  \nonumber\\
  & =
  -
  \frac{2}{\ap}
  \sum_{ \RR=1}^{N-1} \sum_{ \TT= \RR+1}^{N}
  \sum_{i=2}^{D-2} 
  \oint_{Z_{\sN \RR}=0} \frac{d Z_{\sN \RR}}{2\pi i}
  \oint_{Z_{\sN \TT}=0} \frac{d Z_{\sN \TT}}{2\pi i}
  \partial \uL_{\sN \RR (\ne) }^{(+) i} (Z_{\sN \RR} )
  \,
  \partial \uL_{\sN \TT (\ne) }^{(+) i} (Z_{\sN \TT} )
  \,
      \ln \left(
      z(Z_{\sN \RR})  -z(Z_{\sN \TT})
    \right)
    .
  \label{eq:interaction_an_an}
\end{align}

Using the trivial fact that
\begin{equation}
  \uL_{\sN \RR  }^{(+) i} (Z_{\sN \RR} )
  =
  \uL_{\sN \RR (\ne) }^{(+) i} (Z_{\sN \RR} )
  +
  \uL_{\sN \RR (0) }^{(+) i} (Z_{\sN \RR} )
  ,
\end{equation}
we can now assemble all interactions with $r\ne \TT$ given in
eq.s \eqref{eq:interaction_an_a0}.
\eqref{eq:interaction_a0_a0}
and \eqref{eq:interaction_an_an}
under a very simple expression
\begin{align}
  \sum_{\RR \ne \TT}
  &
  \left[
  \Alc_{\sN \RR n} \Alc_{\sN \TT 0}\mbox{ terms }
  +
  \Alc_{\sN \RR 0} \Alc_{\sN \TT 0}\mbox{ terms }
  +
  \Alc_{\sN \RR n} \Alc_{\sN \TT m}\mbox{ terms }
  \right]
  =
  \nonumber\\
  =&
  -
  \frac{2}{\ap}
  \sum_{ \RR=1}^{N-1} \sum_{ \TT= \RR+1}^{N}
  \sum_{i=2}^{D-2} 
  \oint_{Z_{\sN \RR}=0} \frac{d Z_{\sN \RR}}{2\pi i}
  \oint_{Z_{\sN \TT}=0} \frac{d Z_{\sN \TT}}{2\pi i}
  \partial \uL_{\sN \RR }^{(+) i} (Z_{\sN \RR} )
  \,
  \partial \uL_{\sN \TT  }^{(+) i} (Z_{\sN \TT} )
  \,
      \ln \left(
      z(Z_{\sN \RR})  -z(Z_{\sN \TT})
    \right)
    .
  \label{eq:interaction_r_ne_t}
\end{align}
The previous expression strongly suggests that in DDF correlators
there are hidden the Mandelstam maps.
In fact the previous equation \eqref{eq:interaction_r_ne_t} is very
similar to the usual one in eq. \eqref{eq:N Reggeon} up to use of
different maps, i.e eq. \eqref{eq:usual_local_global} for the usual
case  and eq. \eqref{eq:Zr zr}   for the DDF case.
Explicitly we identify
$z(Z_{\sN \RR}) = m_{\sN \RR}(Z_{\sN \RR})$.

Looking to the whole Reggeon we see that the DDF Reggeon
\eqref{eq:Reggeon_rewritten_v1} has further contributions, notably
some self interactions and a normalization term.

We will show in the next section that self interaction terms arise
from the conformal transformation of the \lc SDS vertex (given in
eq. \eqref{eq:SDS_lc_1}) under the
local Mandelstam maps suggested by the previous expression.

We are then left to explain the extra normalization factor in the first line in
eq. \eqref{eq:Reggeon_rewritten_v1}.
We will argue that this is the factor Mandelstam got from the
functional determinant of $2d$ Laplacian and other normalization factors
in path integral.
The reason is simple and is that DDF amplitudes are covariant and the
factors computed by Mandelstam are needed for covariance.

\section{Recovering Mandelstam map 4: Mandelstam from self interactions}
\label{sec:Mandelstam_for_self_interacting_strings}

As explained at the end of the previous section we need to explain the
self interactions.
In particular we would interpret eq. \eqref{eq:Reggeon_rewritten_v1}
as a result of a computation similar to the usual one given in
eq. \eqref{eq:derivation_Reggeon_tree_level}.
Explicitly we would like to show that (with $|x_1|>\dots>|x_N|$)

\begin{align}
  \langle V_N^{(DDF+ghost)}( \{ \xr \})|
  =&
  \cN( \{ \xr \})\,
  \delta\left(
    \sum_{\RR=1}^N \uk_{\sN \RR}^+
    \right)\,
    \,
    \delta\left(
    \sum_{\RR=1}^N \uk_{\sN \RR }^-
    \right)\,
  \nonumber\\
  &
  \langle \tilde 0_\lc |\,
  m_{\sN 1} \circ \widetilde \cS_{\lc \sN 1}(0)
  \dots
  m_{\sN N} \circ \widetilde \cS_{\lc \sN N}(0)\,
  | \tilde 0_\lc\rangle_{\widetilde{UHP}}
  ,
  \label{eq:derivation_DDF_lc_Reggeon_tree_level}
\end{align}
where the basic \lc SDS are given in eq. \eqref{eq:SDS_lc_1},
$ m_{\sN \RR}(0)=\xr$ and
the normalization factor $\cN( \{ \xr \})$ is discussed later.
Notice that we have written $\widetilde{UHP}$ to make clear that the
expectation value is \wrt $\widetilde \uLlc$ fields.

Since the contribution from the $\widetilde \uLlc$ has been considered in the
previous section and is completely analogous to the usual one
since it gives
$\ln \left(z(Z_{\sN \RR})  -z(Z_{\sN \TT}) \right)
=
\ln \left(m_{\sN \RR}(Z_{\sN \RR})  -m_{\sN \TT}(Z_{\sN \TT}) \right)
$,
we would now consider the contributions from self interactions.
In particular we would like to show that self interactions arise from
regularizing the \lc SDS vertex on the previous Riemann surface.
In fact the self interaction is the left over of the regularization of
the non normal ordered \lc SDS vertex
or, more exactly, the effect of a non trivial conformal transformation.

In order to do so we write the SDS vertex for the \lc fields as
\begin{align}
  \widetilde \cS_\lc(\XX)
  &=
  \langle x^i=0, 0_{\Alc^i} |
  :
  e^{
    \frac{i}{\sdap}
   \delta_{i j}\,
    \sum_{n=0}^\infty \frac{\Alc^i_{n}}{n!}
    \partial^n_x {\widetilde \uLlc}^j(\XX, \bar \XX)
  }
  :
  \nonumber\\
  &=
  \langle x^i=0, 0_{\Alc^i} |
  :
  e^{
    -\frac{2}{\ap}
    \delta_{i j}
    \oint_{\ZZ=0,\, |\ZZ|<|\XX|} \frac{d \ZZ}{ 2\pi i}
    \partial_\ZZ \uLlc^i(\ZZ)
     {\widetilde \uLlc}^j(\XX +\ZZ)
  }
  :
  ,
  \label{eq:SDS_lc_1}
\end{align}
and notice that this vertex has very different conformal properties of
the apparently analogous vertex for DDF operators
\begin{align}
  \widetilde \cS_{DDF}(x)
  &=
  \langle x^\mu=0, 0_{\uA^i} |
  :
  e^{
    \frac{i}{\sdap}
   \delta_{i j}\,
    \sum_{n=0}^\infty \frac{\uA^i_{n}}{n!}
        {\widetilde \uA}^j_n(x)
  }
  :\,
  :\,e^{i \uk_{T \mu} {\widetilde \uL}^\mu(x, \bar x)}\,:
  ,
\end{align}
where ${\widetilde \uA}^j_n(x)$ is the DDF operator at $x$ which
contains both creators and annihilators of the covariant field.
The reason is that $\tilde \uA^i_n(u)$ (and $\uA^i_n$) are
good zero dimensional conformal operators,
i.e. primary fields while
$\partial^n_x {\widetilde \uLlc}^j(x, \bar x)$
(i.e. $\widetilde \Alc^i_n$ and $\Alc^i_n$)
are for $n\ne 1$ not primary fields.

This implies that under a conformal transformation
$u=f(U)$ (in the UHP)
generated by an operator $\tilde U_f$
and acting on the tilded fields they transform as
(see appendix \eqref{app:conf_transf_SDS_details})
\begin{align}
  \tilde U_f\,
  &
  \widetilde \cS_{\lc}(\XX)\,
  \tilde U_f^{-1}
  =
  f\circ   \widetilde \cS_{\lc}(\XX)
  \nonumber\\
  =
  &
  \widetilde \cS_{\lc}( f(\XX) )
  \nonumber\\
  &
  \times
  \lim_{\epsilon \rightarrow 0}\,
    e^{
    \frac{2}{\ap^2}
    \oint_{\WW=\epsilon,\,} \frac{d \WW}{ 2\pi i}
    \oint_{\ZZ=0,\, |\ZZ|<\epsilon} \frac{d \ZZ}{ 2\pi i}
    \partial_\WW \uLlc^i(\WW -\epsilon)\,
    \partial_\ZZ \uLlc^i(\ZZ)\,
    \Delta_f(\XX +\WW, \XX +\ZZ)
    }
    ,
    \label{eq:lcSDS_conf_trasformation}
\end{align}
where
\begin{align}
  \Delta_f(\XX +\WW, \XX +\ZZ)
  &
  =
  G_0(f(\XX +\WW), f(\XX+ \ZZ) ) -G_0(\XX +\WW, \XX+ \ZZ)
  \nonumber\\
  =&
  -\hap\,
  \ln\frac{f(\XX +\WW)- f(\XX+ \ZZ) }{\WW -\ZZ}
. 
\end{align}
This transformation has to be compared with
\begin{align}
  \tilde U_f\,
  \widetilde \cS_{DDF}(x)\,
  \tilde U_f^{-1}
  =&
  f\circ   \widetilde \cS_{DDF}(x)
  =
  \left( \frac{d f}{d x} \right)^{
   \oh \uA^i_{0} \uA^i_{0}
  }\,
  \widetilde \cS_{DDF}( f(x) )
  .
\end{align}

We can now apply the previous discussion to the case where
$u = f(U)$ is $u-\xr=f_{\sN \RR}(U_{\sN \RR})$ and $\XX=0$ and compare with
eq. \eqref{eq:Reggeon_rewritten_v1}.
We can then use
$U_{\sN \RR}  = F_{\sN \RR}(u-\xr) = U_{\sN \RR}(u)$
and
$\XX=0 \Rightarrow U_{\sN \RR}=0 $ so that
$u-\xr=0$.
Notice that in the text but not here we wrote
$u=m_{\sN \RR}(U_{\sN \RR})$ instead of $u=u(U_{\sN \RR})$, the reason being
that here is more functional to use $u(U_{\sN \RR})$ since this makes
clear that they are the inverse but in the text we want to stress that
they are the local Mandelstam maps.
If we write eq. \eqref{eq:lcSDS_conf_trasformation} in a more explicit
way we get
\begin{align}
  u \circ   \widetilde \cS_{\lc}(U_{\sN \RR})
  =
  &
  \widetilde \cS_{\lc}( u( U_{\sN \RR}) )
  \nonumber\\
  &
  \times
  \lim_{\epsilon \rightarrow 0}\,
  e^{
    \frac{2}{\ap^2}
    \oint_{\WW=\epsilon,\,} \frac{d \WW}{ 2\pi i}
    \oint_{\ZZ=0,\, |\ZZ|<\epsilon} \frac{d \ZZ}{ 2\pi i}
    \partial_\WW \uLlc^i(\WW-\epsilon)\,
    \partial_\ZZ \uLlc^i(\ZZ)\,
    \Delta_u(\XX +\WW, \XX +\ZZ)
  }
  \nonumber\\
  =
  &
  \widetilde \cS_{\lc}( u( U_{\sN \RR}) )
  \times\,
  e^{+\oh\,
    \Alc_{\sN \RR 0}^i \Alc_{\sN \RR 0}^i\,
    \left. \ln \frac{d u}{d U_{\sN \RR} }\right|_{U_{\sN \RR}=0}
  }
  \nonumber\\
  &
  \times
  e^{
     \frac{-2}{\ap}\,
    \sum_{n=1}^\infty\,
    \Alc_{\sN \RR 0}^i \Alc_{\sN \RR n}^i\,
       \frac{1}{n!}
    \left. \del_{U}^{n}\Delta_u(U,0)\right|_{U=0}
  }
  \nonumber\\
  &
  \times
  e^{-\frac{1}{\ap} \sum_{n, m=1}^\infty\,
    \Alc_{\sN \RR n}^i \Alc_{\sN \RR m}^i\,
    \frac{1}{n! m!}
    \left. \del_{U_1}^{n} \del_{U_2}^{m} \Delta_u(U_1, U_2)\right|_{U_{1,2}=0}
  }
  ,
  \label{eq:lcSDS_conf_trasformation_expanded}
\end{align}

We notice immediately that the expression in
eq. \eqref{eq:Reggeon_rewritten_v1} is in term of the inverse function
$U=F(u)$.
In appendix \ref{app:conf_transf_SDS_details}
we give a proof of the required identity
\begin{align}
  \lim_{U_i \rightarrow 0}
  \frac{1}{m!n!}\,
  \frac{-2}{\ap}\,
  \del_{U_1}^{m}\del_{U_2}^{n}\Delta_f(U_1,U_2)
  =&
  \frac{1}{m n}
  \oint_{z_1=0}\frac{d z_1}{2\pi i}\,
  \oint_{z_2=0,|z_1|>|z_2|}\frac{d z_2}{2\pi i}\,
  \frac{1}{(z_1-z_2)^2}\frac{1}{[F(z_1)]^{m}[F(z_2)]^{n}}
  ,
\label{eq:toproveSDSselfint_maintext}
\end{align}
for $m+n\ge 1$ which allows to perform the desired map.

\section{Recovering Mandelstam map 5: the normalization factor
  when on shell}
\label{sec:Nomalization_factor_vs_Mandelstam}

We would now discuss the normalization factor
\begin{align}
  \cN(\{\xr\})
  =&
  \prod_{\RR=1}^{N}
  \Biggl\{
  \left[
    \frac{d \Zlc{ \RR}(z)}{d z}
    \Biggr|_{z=\xr}
    \right]^{
    + \ap m_{T \sN \RR}^2
  }
  \Biggr\}
  (x_1-x_2)\,(x_1-x_N)\,(x_2-x_N)
  ,
\end{align}
in eq. \eqref{eq:Reggeon_rewritten_v1}.

We would like to argue that in {\sl critical dimension} and {\sl on shell}
when $\ap m_{T \sN \RR}^2=-1$ is {\sl essentially} the factor
computed by Mandelstam in path integral approach
which comes 
from the functional determinant of the $2d$ Laplacian and the Jacobian
which arises from the change of coordinates from $\tau_E + i \sigma$
to upper plane ones (\cite{Mandelstam:1973jk}
and \cite{Green:1987mn} vol. 2
and \cite{Baba:2009ns} for a more recent approach).

The reason is simple and is that {\sl on shell} DDF amplitudes in {\sl
  critical dimension} are covariant and the normalization factor
$\cN(\{\xr\})$ is needed to get a covariant amplitude exactly as the
factor computed by Mandelstam is needed for covariance.

There is also another reason to stress on shell.
Mandelstam results are in an Hamiltonian framework where there is
not \lc energy conservation unless the consider an infinite period of
evolution, i.e. the $S$ matrix.
When this happens we have $ \sum_{\RR} p^-_{\RR}=0$
so the states are on shell.

This is also the reason why we have to add by hand the
$    \delta\left(
    \sum_{\RR=1}^N \uk_{\sN \RR }^-
    \right)\,
$
in \eqref{eq:derivation_DDF_lc_Reggeon_tree_level}.

On the other side
$ \delta\left(    \sum_{\RR=1}^N \uk_{\sN \RR }^+ \right)$
is implicit in Mandelstam work because of the conservation of the
strip width imposed by hand.

We would now like to comment on the adverb ``essentially''.
Mandelstam computations are performed always choosing one vertex in
the upper half plane at $x=\infty$.
The way we have performed the computation with the DDF vertexes does
not allow to send any coordinate to $\infty$.
On the other hand on shell we know that the amplitude can be written
in terms of anharmonic ratios which allow to take one coordinate to
infinity in a smooth way.
Therefore we take one coordinate to infinity and we drop any divergent
factor in the Reggeon since it must be canceled when expanding it for
computing an explicit amplitude.
This argument is indirect but can be checked directly in the $N=3$
case.
In particular 
we can compare in details the integrals,
i.e. the Neumann coefficients in the case $N=3$ and find a perfect match.
Details are given in appendix \ref{app:Reggeon_N=3_comparison}.





\section{Going off shell: a difference}
\label{sec:Mandelstam_off_shell}

Finally we can consider what happens {\sl off shell}.
Off shell there is a difference.
In this section we keep the discussion at the conceptual level and we
write the details
in appendix \ref{app:Reggeon_N=3_comparison}.

Before we compare we need to explain what we mean to go off shell in
our case.

The simplest way is to let the tachyon masses to be free parameters.
In doing so we get amplitudes which depend on the Koba-Nielsen
variables and are not unique.
This is to be expected since off shell physics is not unique.
On the other hand if we want an off shell extension which is well
defined and consistent we can consider the Witten three vertex for the
$N=3$ amplitudes and then build the others from it.

Let us consider the Witten vertex.
Instead of the usual basis of covariant states of ghost number $-1$
we can use a basis formed by DDF, improved Brower states along with the
dual to the improved Brower states.
These states do not obviously include the zero momentum states.
This basis is however interesting since on shell only DDF states are
physical and non null while improved Brower states are BRST exact, i.e
physical and null.
In order to compare with the \lc we can restrict our attention to DDF states.
We notice that the DDF states and operators are primary even when the tachyon
mass is a free parameter.
In facts the framed DDF operators $\uA^i_n(x)$ are conformal operators
of dimension zero.
It follows that dimension of an off shell DDF state is equal to the
dimension of the associated tachyon vertex, i.e. $\ap \uk_T^2$.
The Witten vertex is then simply
\begin{align}
  \langle W| \Bigg|_{DDF}
  =&
  \left( \frac{8}{3} \right)^{\ap \uk_{T \sN 1}^2}\,
  \left( \frac{2}{3} \right)^{\ap \uk_{T \sN 2}^2}\,
  \left( \frac{8}{3} \right)^{\ap \uk_{T \sN 3}^2}\,
  \langle{V}^{(DDF+ghost)}_{N=3}(\{ x_3=-\sqrt{3}, x_2=0, x_1=\sqrt{3} \})|
  .
\end{align}
Because DDF states are primary operators this discussion
(see \cite{Erler:2020beb} and also \cite{Baba:2009kr,Baba:2009ns})
can be extended to any
consistent string field theory whose maps for the cubic order are
$u=f_{3 \sN \RR}(U_{\sN \RR})$ as
\begin{align}
  \langle V_3| \Bigg|_{DDF}
  =&
  \left( f_{3 \sN 1}'(0) \right)^{\ap \uk_{T \sN 1}^2}\,
  \left( f_{3 \sN 2}'(0) \right)^{\ap \uk_{T \sN 2}^2}\,
  \left( f_{3 \sN 3}'(0) \right)^{\ap \uk_{T \sN 3}^2}\,
  \nonumber\\
  &
  \langle{V}^{(DDF+ghost)}_{N=3}(\{ x_3=f_{3 \sN 3}(0), x_2=f_{3 \sN 2}(0), x_1=f_{3 \sN 1}(0) \})|
  .
  \label{eq:V3_generic_SFT}
\end{align}

Notice that conceptually there is also a difference in the approaches.
Here we start from off shell DDF states as discussed in \cite{Biswas:2024unn}
and we use them as (a subset of) a basis of string states
and then we  compute the off shell amplitude for $N=3$ states
according to string field theory recipes.
From this amplitude we read the Mandelstam maps which match the result
from first quantized sting theory because of
eq. \eqref{eq:V3_generic_SFT}.
These maps are intrinsic to the DDF states and not imposed or chosen
and the eventual singularities in the local half disk are also intrinsic.
Differently in \cite{Baba:2009kr,Baba:2009ns} they assume a given
expression for the Mandelstam maps but then the expressions for the
Mandelstam maps do not match.
Comparison with \cite{Erler:2020beb} is less straightforward since in
this paper they integrate out the Brower states (or at least states
which should correspond to Brower states even if their expression does
not match the usual expression for Brower states see eq. 3.80 of \cite{Erler:2020beb}).

Now we can compare with \lc formalism.
First of all we notice that
Mandelstam is working in Hamiltonian formalism and ``off
shell'' means evolution for a finite time which implies that \lc
energy is not conserved, i.e. there is {\sl no}
$\delta\left(\sum_{\RR=1}^N \uk_{\sN \RR }^- \right) $.

When considering finite time intervals
Mandelstam Reggeon vertex acquires a dependence on
the interaction times, for the $N=3$ only one interaction time $\tau_0$.
The precise form of $\tau_0$ seems to be fundamental to prove the \lc
factorization of the $N=4$ amplitudes (see paragraph 7 of
\cite{Mandelstam:1973jk}).

We cannot then compare directly with Mandelstam work when off shell
because we are computing an off shell Green function and he is
computing a matrix $U(\tau_f, \tau_i)$ element.
We can however compare with the formulation of \lc String Field Theory proposed
in \cite{Kaku:1974zz,Cremmer:1974ej,Cremmer:1974jq}.
In this paper Kaku and Kikkawa argue that the very same vertex
of Mandelstam extended off shell can be used
for building the $3$ string functional quantum {\sl Hamiltonian}
in interaction picture, i.e $H_{I\,3}(\tau)$, as
\begin{align}
  H_{I\,3}(\tau)
  =&
  \sum_{\phi_1,\phi_2,\phi_3}\,
  \left(
  h_{3\, \phi_1,\phi_2,\phi_3}\,
  A^\dagger_{I\,\phi_1}(\tau)\,
  A_{I\,\phi_2}(\tau)\,A_{I\,\phi_3}(\tau)\,
  +h.c.
  \right)
  \nonumber\\
  =&
  \sum_{\phi_1,\phi_2,\phi_3}\,
  \left(
  h_{3\, \phi_1,\phi_2,\phi_3}\,
  e^{-i \tau ( \hat P_{\sN 1 +}  -\hat P_{\sN 2 +}  -\hat P_{\sN 3 +} ) }
  A^\dagger_{\phi_1}\,
  A_{\phi_2}\,A_{\phi_3}\,
  +h.c.
  \right)
  .
  \label{eq:H3_kaku_kikkawa}
\end{align}
In this expression $\phi_\RR$ are ``particles'', i.e.
second quantized \lc string states.
Each $\phi$ is in correspondence with a first quantized \lc string and
therefore is labeled by $| k_i,\, k_-,\, \{ N_{i n} \} \rangle$.
The creator $A_\phi^\dagger$
(which is in the Schroedinger picture
while
$A_{I\, \phi}^\dagger(\tau)$ is in the interaction picture
) must not be confused with $\uA^i_{n}$ since
the latter is a first quantized DDF operator while the second is a
second quantized creator which creates a first quantized string
$\prod_{i,\, n} \left( \Alc^i_{-n} \right)^{N_{i n}}\,  | k_i,\, k_-
\rangle$.
It creates a second quantized string $A_\phi^\dagger |\Omega\rangle$
{}\, \footnote{
In particular $\prod_{\RR=1}^M A_{\phi_{\sN \RR}}^\dagger
|\Omega\rangle$ is a multi string state composed by $M$ strings.
}
which is eigenstate of the second quantized free Hamiltonian $H_2$ with
eigenvalue $P_+$, i.e.
\begin{align}
  H_{2}\, A_\phi^\dagger |\Omega\rangle
  =&
  \hat P_{+}\, A_\phi^\dagger |\Omega\rangle,
  ~~~~
  \hat P_{+}
  = \frac{\sum_i k_i^2 + \sum_{i,n} N_{i n} -1}{2 k_-}
  .
\end{align}

As a consequence the coefficient $h_{3\, \phi_1,\phi_2,\phi_3}$ can be written as
\begin{align}
    h_{3\, \phi_1,\phi_2,\phi_3}\,
    = c\left(
    \tau_0,\,
    \left\{ k_{\sN \RR\, i},\, k_{\sN \RR\, -},\, \{ N_{\RR\,i n} \} \right\}_{\RR=1,2,3}
    \right)\,
    \delta^{D-2}\left(\sum_{\RR=1}^N \uk_{\sN \RR i} \right)\,
    \delta\left(\sum_{\RR=1}^N \uk_{\sN \RR -} \right)
    ,
\end{align}
where Kaku and Kikkawa use explicitly Mandelstam vertex given in
eq. \eqref{eq:V3_Mandelstam} as
\begin{align}
    h_{3\, \phi_1,\phi_2,\phi_3}\,
    =&
    \langle V_3^{(M)}(0) |\,
    \left| k_{\sN 1 \, i},\, k_{\sN 1\, -},\, \{ N_{\sN 1\,i n} \}
    \right\rangle
    \otimes
    \left| k_{\sN 2 \, i},\, k_{\sN 2\, -},\, \{ N_{\sN 2\,i n} \}
    \right\rangle
    \otimes
    \left| k_{\sN 3 \, i},\, k_{\sN 3\, -},\, \{ N_{\sN 3\,i n} \}
    \right\rangle
    \nonumber\\
    &\times
        \delta\left(\sum_{\RR=1}^N \uk_{\sN \RR -} \right)
    .
\end{align}
The use of Mandelstam vertex extended off shell was further
researched in \cite{Hopkinson:1975pm} without a clear final answer
whether it is fully correct.

To compare with the modem point of view according to which  tree level
string field theory is classical we have to ``de-quantize''
eq. \eqref{eq:H3_kaku_kikkawa}
and make the quantum operators $A_{I \phi}(\tau)$ classical modes
of a classical \lc field
$\Phi_{\{N_{i n} \} }(k_i,\,k_-,\, k_+)$
and then integrate over the \lc time $\tau=X^+$
as $\int_{-\infty}^{+\infty} d\tau $
(see also \cite{Baba:2009kr,Baba:2009ns})
to get finally the classical action
\begin{align}
  S_{3}^{(lc)}
  =
  \int\, \prod_{\RR=1}^3 d^D k_{\sN \RR}
  &
  \sum_{ \{ N_{\sN \RR \,i n} \} }\,
  \Bigg(
  \langle V_3^{(M)}(0) |\,
    \otimes_{\RR=1}^3
    \left| k_{\sN \RR \, i},\, k_{\sN \RR\, -},\, \{ N_{\sN \RR\,i n} \}
    \right\rangle
    \,\,
    \delta^D( k_{\sN 1 }  -k_{\sN 2}  -k_{\sN 3 } )
    \nonumber\\
    &
  \Phi^*_{\{N_{\sN 1 i n} \} }(k_{\sN 1 \mu})\,
  \Phi_{\{N_{\sN 2 i n} \} }(k_{\sN 2 \mu})\,
  \Phi_{\{N_{\sN 3 i n} \} }(k_{\sN 3 \mu})\,
  +h.c.
  \Bigg)
  .
  \label{eq:classical_S3_kaku_kikkawa}
\end{align}

This classical action can then be compared with the classical action
obtained from  \eqref{eq:V3_generic_SFT}.
Kaku and Kikkawa use the vertex $\langle V_3^{(M)}(0) |$ which
is given in term of the Neumann coefficients $\bar N$ (see \eqref{eq:Mandelstam_N_bar_N_f_etc}).
These Neumann coefficients $\bar N$ have an explicit $\tau_0$
dependence which is well defined function of $\sdap \ukp_{\sN r}$ (see
eq. \eqref{eq:Mandelstam_N_bar_N_f_etc}).
Now the classical action obtained from \eqref{eq:V3_generic_SFT}
has Neumann coefficients which do not depend on $\sdap \ukp_{\sN r}$
unless the maps $u=f_{3 \sN \RR}(U_{\sN \RR})$ are chosen to depend
explicitly on $\sdap \ukp_{\sN r}$ (see
eq. \eqref{eq:comparison_bar_N_Neumann_DDF}) .
This means that the results do not agree.

Would they use $\langle V_3^{(M)}(\tau_0) |$ the their vertex would
contain the Neumann coefficients $N$ and this would match our results
in the limit $x_3\rightarrow \infty$
despite the fact that Mandelstam uses $\langle V_3^{(M)}(0) |$ for proving the $N=4$ factorization.

The reason of the difference cannot to be traced to the
existence of improved Brower states which are null, i.e. BRST exact
only on shell but can enter the internal off shell leg since we are
considering the $N=3$ Green function.


\section*{Acknowledgments}
We would like to thank Raffaele Marotta for discussions.  This
research is partially supported by the MUR PRIN contract 2020KR4KN2
“String Theory as a bridge between Gauge Theories and Quantum Gravity”
and by the INFN project ST\&FI “String Theory \& Fundamental
Interactions”.


\appendix

\section{Details on conformal transformation of \lc SDS}
\label{app:conf_transf_SDS_details}

We want to consider the transformation of the \lc SDS under a generic
conformal transformation.
To this purpose we consider just one \lc direction
since they are independent.
The non trivial transformation is
the origin of the self-interaction integral appearing in the SDS
correlator generating function.

We proceed naively and then we justify in a more rigorous way the
result.
We use capital coordinates like $\XX, \UU$ since we think they are
local coordinates.  
We start from the point splitted version of the SDS vertex
\begin{align}
[\Tilde{S}(\XX)]_{p.s}
=
\langle 0; 0_a|\,
\exp\Bigg[
  &\frac{2}{\ap}
  \Big(
  \oint_{\ZZ=0; |\ZZ|<\epsilon}\frac{d \ZZ}{2\pi i}\,
  \del L^{(+)}(\ZZ) \Tilde{L}^{(-)}(\XX +\ZZ)
  \nonumber \\
  &+ \oint_{\WW= \epsilon}\frac{d \WW}{2\pi i}\,
  \del L^{(+)}(\WW -\epsilon) \Tilde{L}^{(+)}(\XX +\WW)
  \Big)
  \Bigg]
,
\label{eq:SDSps}
\end{align}
and then we normal order it as
\begin{align}
  & [\Tilde{S}(\XX)]_{p.s}
  =
  \nonumber\\
  =&
  \langle 0; 0_a|\,
:..:
\exp
\Bigg\{
-\frac{1}{2}
\frac{4}{ {\ap}^2 }
\oint_{\ZZ=0} \frac{d \ZZ}{2\pi i}\,
\oint_{\WW=\epsilon} \frac{d \WW}{2\pi i}\,
\left[
  \del L^{(+)}(\ZZ) \Tilde{L}^{(-)}(\XX+\ZZ),
  \del L^{(+)}(\WW-\epsilon) \Tilde{L}^{(+)}(\XX+\WW)
  \right]
  \Bigg\} 
\nonumber \\
=&
\langle 0; 0_a|\,
:..:
\exp\left[
  \frac{1}{2}
  \frac{4}{ {\ap}^2 }
  \oint_{\ZZ=0} \frac{d \ZZ}{2\pi i}\,
  \oint_{\WW=\epsilon} \frac{d \WW}{2\pi i}\,
  \del L^{(+)}(\ZZ) \del L^{(+)}(\WW-\epsilon)\, G_0(\XX+ \WW, \XX +\ZZ)
  \right],
\end{align}
where we have used
$[\Tilde{L}^{(+)}(\WW), \Tilde{L}^{(-)}(\ZZ)] =G_0(\WW,\ZZ) =G_0(\WW-\ZZ)$
in obtaining the last line.

The regularized SDS vertex, i.e. the normal ordered one
is then obtained by removing the UV
divergent extra
factor above and taking the limit $\epsilon \rightarrow 0$,
\begin{equation}
  [\Tilde{S}(\XX)]_{reg} =
  :\Tilde{S}(\XX): =
  \lim_{\epsilon  \rightarrow 0}
  [\Tilde{S}(\XX)]_{p.s}\,
  \mathcal{N}_0(\XX, \epsilon)
  ,
\label{eq:sdsreg}
\end{equation}
where
\begin{equation}
\mathcal{N}_0(\XX, \epsilon) =
\exp\left[
  -\frac{1}{2}
  \frac{4}{ {\ap}^2 }
  \oint_{\ZZ=0}\frac{d \ZZ}{2\pi i}\,
  \oint_{\WW=\epsilon}\frac{d \WW}{2\pi i} \,
  \del L^{(+)}(\ZZ)\, \del L^{(+)}(\WW-\epsilon)\, G_0(\XX +\WW, \XX +\ZZ)
  \right]
  .
\end{equation}
Notice that $\mathcal{N}_0$ is operatorial and must be written after 
$[\Tilde{S}(\XX)]_{p.s}$ which contains a $\langle 0; 0_a|$ on which
$\mathcal{N}_0$ acts.

We now apply the conformal transformation $u=f(U)$ generated by
$\tilde U_f$ on the chiral tilded field $\tilde L(\UU)$.
We proceed naively and apply the transformation on the regularized
expression not taking into account that $\tilde L(\UU)$ is point
splitted and not really a primary field.
We get an expression that we verify in a different way.

We find naively
\begin{align}
\tilde U_f\,
[\Tilde{S}(\XX)]_{p.s}\,
\tilde U_f^{-1}
=
\langle 0; 0_a|\,
\exp
\Bigg[
  \frac{2}{\ap}
  \Big(
  \oint_{\ZZ=0}\frac{d \ZZ}{2\pi i}\,
  \del L^{(+)}(\ZZ)\, \Tilde{L}^{(-)}( f(\XX +\ZZ) )
  \nonumber \\
  + \oint_{\WW=\epsilon}\frac{d \WW}{2\pi i}\,
  \del L^{(+)}(\WW-\epsilon) \Tilde{L}^{(+)}( f(\XX +\WW) )
  \Big)
  \Bigg]
,
\label{eq:f_on_SDSps}
\end{align}
then we normal order and consider the regularized version
\begin{align}
\tilde U_f\,
 &
[\Tilde{S}(\XX)]_{reg}\,
\tilde U_f^{-1}
=
f \circ [\Tilde{S}(\XX)]_{reg}
=
\nonumber\\
=&
\langle 0; 0_a|\,
:\,
\exp\Bigg[\frac{2}{\ap}
  \Big(
  \oint_{\ZZ=0}\frac{d \ZZ}{2\pi i}\,
  \del L^{(+)}(\ZZ) \Tilde{L}( f(\XX+\ZZ) )
  \Big)
  \Bigg]\,
:
\nonumber\\
\times
&
\lim_{\epsilon \rightarrow 0}
\exp\left[
  \frac{1}{2}
  \frac{4}{ {\ap}^2 }
  \oint_{\ZZ=0}\frac{d \ZZ}{2\pi i}\,
  \oint_{\WW=\epsilon}\frac{d \WW}{2\pi i} \,
  \del L^{(+)}(\ZZ)\, \del L^{(+)}(\WW-\epsilon)\,
  \Delta(\XX +\WW, \XX +\ZZ)
  \right]
,
\end{align}
where the conformal transformation produces the non trivial
(operatorial) factor
\begin{align}
  \cN(\XX)
  =
\lim_{\epsilon \rightarrow 0}
\exp\left[
  \frac{1}{2}
  \frac{4}{ {\ap}^2 }
  \oint_{\ZZ=0}\frac{d \ZZ}{2\pi i}\,
  \oint_{\WW=\epsilon}\frac{d \WW}{2\pi i} \,
  \del L^{(+)}(\ZZ)\, \del L^{(+)}(\WW-\epsilon)\,
  \Delta(\XX +\WW, \XX +\ZZ)
  \right]
,
\label{eq:f_circ_SDS}
\end{align}
in which we have defined
\begin{align}
  \Delta(\XX +\WW, \XX +\ZZ)
  =&
  G_0(f(\XX +\WW), f(\XX+ \ZZ) ) -G_0(\XX +\WW, \XX+ \ZZ)
  \nonumber\\
  =&
  -\hap
  \ln\left( \frac{f(\XX +\WW) - f(\XX+ \ZZ) }{ \WW -\ZZ} \right),
\end{align}
which is well behaved as $\WW \rightarrow \ZZ$
($\epsilon \rightarrow 0$) since the UV singularity is
canceled by the $G_0(\XX +\WW, \XX+ \ZZ)$ subtraction.

We notice that we can use eq. \eqref{eq:lambda_to_alpha_rewriting}
to rewrite the previous transformation in a
coherent state approach as
\begin{align}
  f
  \circ
  &
  :\,
  e^{
    i \shap
    \left[
      \lambda_0 \tilde L(X)
    + 
    \sum_{n=1}^\infty \frac{\lambda_n}{(n-1)!}\, \partial_X^n \tilde L(X)
    \right]
    }\, :\,\,
=
\nonumber\\
=&
  :\,
  e^{
    i \shap
    \left[
      \lambda_0 \tilde L( f(X) )
    + 
    \sum_{n=1}^\infty \frac{\lambda_n}{(n-1)!}\, \partial_X^n \tilde L( f(X) )
    \right]
    }\, :
\nonumber\\
\times
&
( f'(X) )^{\oh \lambda_0^2}\,
e^{
  -\frac{2}{\ap}
  \lambda_0
   \sum_{n=1}^\infty \frac{\lambda_n}{(n-1)!}\, \partial_X^n \Delta(X,Y)|_{Y=0}
}\,
e^{
  -\frac{1}{\ap}
   \sum_{n=1}^\infty \frac{\lambda_n}{(n-1)!}\,
   \sum_{m=1}^\infty \frac{\lambda_m}{(m-1)!}\,
  \partial_X^n \partial_Y^m\Delta(X,Y)|_{Y=0}
}
.
\label{eq:f_circ_SDS_coh_form}
\end{align}

\subsection{Checking the conformal transformation: an example}

The goal of this section is to show that the regularization factor
$\mathcal{N}(x)$ derived above is in fact the correct term despite the
naive approach.
We consider a simple operator $:(\del L)^2:(X)$ with $|U_1|> |U_2|$
\begin{align}
&:\, (\del L)^2\, :\,(X)
=
\lim_{U_a\rightarrow X}
\left[\del L(U_1)\, \del L(U_2) - \del_{U_1}\del_{U_2} G_0(U_1,U_2)\right]
\nonumber \\
\& &\,\,
f \circ (\del L(U_1)\,\del L(U_1))
=
  f'(U_1)\, \del_{U_1}L(f(U_1))\, f'(U_2)\, \del_{U_2}L(f(U_2))\, 
\nonumber \\
\implies &
f \circ :\, (\del L)^2\,:\, (X)
=
\lim_{U_i \rightarrow X}
\Big\{
  \del_{f(U_1)}L(f(U_1))\, \del_{f(U_2)}L(f(U_2))\, 
  +
  \del_{U_1}\del_{U_2} \left[  G_0(f(U_1),f(U_2)) -G_0(U_1,U_2) \right]
  \Big\}
,
\label{eq:dL2eg}
\end{align}
which reproduces the previous naive result.

It is then immediate to generalize the previous example to
$:(\del L)^n\, (\del L)^m:(X)$ by simply taking derivatives \wrt $U_1$
and $U_2$.

For the more general case 
$:\prod_{a=1}^M (\del L)^{n_a}:(X)$ we can proceed as before starting
from
\begin{align}
  \lim_{U_a\rightarrow X}
  \left[
    \del^{n_1} L(U_1)\, \del^{n_2} L(U_2)\dots \del^{n_M} L(U_M)
    - \sum \mbox {possible contractions}
    \right]
  ,
\end{align}
with $|U_1| >  |U_2| > \dots |U_M|$.


\subsection{Comparison with self-interaction integral - generic proof}

In the main text in order to match the result from the conformal
transformation of the \lc SDS under a map $u=f(U)$
with the result from the DDF computation we have to rewrite the result
for the conformal transformation in term of the inverse map $U=F(u)$.

Concretely, we need to prove the following generic relation,
\begin{align}
  \lim_{u_i \rightarrow 0}
  \frac{1}{m!n!}\,
  \frac{-2}{\ap}
  \del_{U_1}^{m+1}\del_{U_2}^{n+1}\Delta_f(U_1 ,U_2)
  =&
  \oint_{z_1=0}\frac{d z_1}{2\pi i}\,
  \oint_{z_2=0,|z_1|>|z_2|}\frac{d z_2}{2\pi i}\,
  \frac{1}{(z_1-z_2)^2}\frac{1}{[F(z_1)]^{m+1}[F(z_2)]^{n+1}},
\label{eq:toproveSDSselfint}
\end{align}
for an arbitrary holomorphic function $u=f(U)$ such that
$f(0)=0 \implies F(0) = 0 $.
We work with the \lhs of \eqref{eq:toproveSDSselfint} and also resolve the limit in a rigorous way as \footnote{It is understood that the limit $\eps \rightarrow 0$ is taken after the substitution $u_2=0, u_1 = \eps$.},
\begin{align}
\lim_{\eps \rightarrow 0} 
\left\{
\frac{1}{m!n!}
\del^m_{U_1}\del^n_{U_2}\,
\left[\frac{f'(U_1)\,f'(U_2)}{(f(U_1)-f(U_2))^2}
  -\frac{1}{(U_1-U_2)^2}\right]
\right\}\Bigg\lvert_{U_1=\eps,~U_2=0} 
\nonumber \\
=
\lim_{\eps \rightarrow 0}
\left\{
\oint_{U_1=\epsilon}\frac{d U_1}{2\pi i}\,
\oint_{U_2=0}\frac{d U_2}{2\pi i}\,
\left[
\frac{f'(U_1)\,f'(U_2)}{(f(U_1)-f(U_2))^2}
-\frac{1}{(U_1-U_2)^2}
\right]
\frac{1}{(U_1 -\epsilon)^{m+1} U_2^{n+1}}
\right\}
\nonumber \\
=
\lim_{\eps \rightarrow 0}
\Bigg\{
\oint_{z_1=f(\epsilon)}\frac{d z_1}{2\pi i}\,
\oint_{z_2=0,|z_1|>|z_2|}\frac{d z_2}{2\pi i}\,
\frac{1}{(z_1-z_2)^2}\frac{1}{[F(z_1)-\epsilon]^{m+1} [F(z_2)]^{n+1}}
\nonumber \\
-
\oint_{U_1=\eps}\frac{d U_1}{2\pi i}\,
\oint_{U_2=0}\frac{d U_2}{2\pi i}\,
\frac{1}{(U_1-U_2)^2}
\frac{1}{(U_1-\eps)^{m+1} U_2^{n+1}}
\Bigg\}
,
 \label{eq:proof}
\end{align}
where $z_i = f(U_i)$. 

Although the individual $\epsilon$ contributions from the two terms in
braces diverge, they exactly cancel each other in the combination in
which they appear in \eqref{eq:proof}. Thus, we can {\sl naively }
take the limit inside and get the self interaction,
\begin{align}
  = \oint_{z_1=0}\frac{d z_1}{2\pi i}\,
  \oint_{z_2=0,|z_1|>|z_2|}\frac{d z_2}{2\pi i}\,
  \frac{1}{(z_1-z_2)^2}\frac{1}{[F(z_1)]^{m+1} [F(z_2)]^{n+1}} .
\end{align}
We note that in writing the last equation, we have used the fact that
the integral
\begin{align}
\oint_{U_1=0}\frac{d U_1}{2\pi i}\, \oint_{U_2=0}\frac{d U_2}{2\pi
  i}\, \frac{1}{(U_1-U_2)^2} \frac{1}{(U_1)^{m+1} U_2^{n+1}}
\end{align}
is zero based on the simple observation that the integrand is
holomorphic for $U_2 \in B_\eps(0)$ (or equivalently for
$|U_2|<|U_1|$).


Since the previous discussion is based on a not rigorous argument
we demonstrate the claim for the lowest order terms.
Explicitly, we compute the following term arising from the regularized SDS vertex,
\eq{
  \lim_{U_i\rightarrow 0}\,
    \frac{-2}{\ap}\,
\del_{U_1}\del_{U_2}\Delta_f(U_1,U_2)
= \frac{c_1 c_3 - c_2^2}{c_1^2},
\label{eq:egm1n1SDSreg}
}
where, $f(U) = \sum_{n=1}^\infty c_n U^n$ is a general holomorphic
function (the inverse of the analytic map \eqref{eq:Zr zr} in our
specific case).
We compare this with the corresponding term obtained from the self-interaction integral to get
\begin{align}
  \oint_{z_1=0}\frac{d z_1}{2\pi i}\,
  \oint_{z_2=0,|z_1|>|z_2|}\frac{d z_2}{2\pi i}\,
  \frac{1}{(z_1-z_2)^2}\frac{1}{[F(z_1)] [F(z_2)]}
  \nonumber \\
  = \sum_{k=1}^\infty
  k\,
  \oint_{z_1=0}\frac{d z_1}{2\pi i}\, \frac{1}{[F(z_1)] z_1^{k+1}}\,
  \oint_{z_2=0}\frac{d z_2}{2\pi i}\,  \frac{z_2^{k-1}}{F(z_2)]}
 . 
\end{align}
We can now write the series expansion of the inverse starting from,
\begin{align}
  w = f(z) =c_1 z + c_2 z^2 + ...
  \nonumber \\
  \implies
  z =  \frac{1}{c_1}\left(w - c_2 z^2 - c_3 z^3 - ..\right)
  \nonumber \\
  \implies
  z = F(w) =\frac{w}{c_1} - \left(\frac{c_2}{c_1^3}w^2 + 2w z^2 \frac{c_2^2}{c_1^3} + \mathcal{O}(w^4)\right) - \left(\frac{c_3}{c_1^4}w^3 + \mathcal{O}(w^4)\right) \nonumber \\
  \implies
  z = \frac{w}{c_1} - \left(\frac{c_2}{c_1^3}\right)w^2 +
  \left(\frac{2c_2^2 - c_1 c_3}{c_1^5}\right)w^3 + \mathcal{O}(w^4)
  ,
\end{align}
and inserting it into the previous series we obtain
\begin{align}
  \sum_{k=1}^\infty
  &
  k\,
  \oint_{z_1=0}\frac{d z_1}{2\pi i}\,  \frac{1}{[F(z_1)]z_1^{k+1}}\,
  \oint_{z_2=0}\frac{d z_2}{2\pi i}\, \frac{z_2^{k-1}}{[F(z_2)]}
  \nonumber\\
  &= \oint_{z_1=0}\frac{d z_1}{2\pi i}\,
  \frac{1}{[F(z_1)]z_1^{2}}\,
  \frac{1}{c_1}
\nonumber\\
&=
\frac{1}{c_1}\, \left(\frac{c_1 c_3 - c_2^2}{c_1}\right)
,
\end{align}
which matches exactly with \eqref{eq:egm1n1SDSreg}.

\section{Details on the comparison with Mandelstam result for $N=3$}
\label{app:Reggeon_N=3_comparison}

We would now give the details of the comparison with Mandelstam's
result for $N=3$.

Mandelstam writes the $N=3$ Reggeon with interaction at $\tau=0$
$|V_3^{(M)}(0) \rangle$ (see \cite{Green:1987mn} vol. 2 section 11.4.5 where the
result is clearer stated than in the original paper) as the time
translated of the corresponding Reggeon computed with the path
integral $|V_3^{(M)} (\tau_0) \rangle$ which comes naturally (with his
conventions) with the interaction at time $\tau_0$ as
\begin{align}
  |V_3^{(M)}(0) \rangle
  =&
  e^{ \oh \tau_0 \sum_{\RR=1}^3 \hat P^-_{\RR} }
  |V_3^{(M)} (\tau_0) \rangle
  \nonumber\\
  =&
  e^{ \oh \tau_0 \sum_{\RR=1}^3 \hat P^-_{\RR} }
  e^{
    \sum_{\RR=1}^3
    N^{\sN \RR}_{m}
    \alpha^i_{\sN \RR -m}\, {\cal P}^i  
    +
    \oh
    \sum_{\sN \RR, \sN \SS=1}^3
    N^{\RR \SS}_{m}
    \alpha^i_{\sN \RR -m}\, \alpha^i_{\sN \SS -n}   
  }
  |0\rangle\,
  \delta^{D-2}(\sum_{r=1}^3 p^i_{\sN \RR})
  \nonumber\\
  =&
  e^{
    -\tau_0 \sum_{\RR=1}^3 \frac{1}{\alpha_{ \RR} }
    +\oh \tau_0
    \sum_{\RR=1}^3 \frac{ p^i_{\sN \RR} p^i_{\sN \RR} }{\alpha_{\sN \RR}}
  }
  e^{
    \sum_{\RR=1}^3
    \bar N^{\RR}_{m}
    \alpha^i_{\sN \RR -m}\, {\cal P}^i  
    +
    \oh
    \sum_{\RR, \SS=1}^3
    \bar N^{\RR \SS}_{m}
    \alpha^i_{\sN \RR -m}\, \alpha^i_{\sN \SS -n}   
  }
  |0\rangle\,
  \delta^{D-2}(\sum_{r=1}^3 p^i_{\sN \RR})  
  ,
  \label{eq:V3_Mandelstam}
\end{align}
where $\hat P^-_{\RR}$ is the \lc Hamiltonian.
The vertexes are obtained in an Hamiltonian framework where there is
not \lc energy conservation unless the consider an infinite period of
evolution, i.e. the $S$ matrix.
When this happens we have $ \sum_{\RR} \hat P^-_{\RR}=0$
so the states are on shell
and the two vertices give the same $S$ matrix.

\begin{figure}[!hbt]
  \centering
  \def\svgwidth{200pt}
  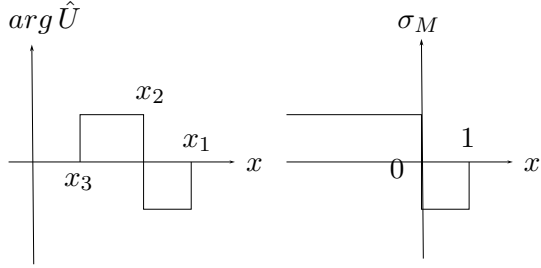
  \caption{The phases of $\hat U$ and the imaginary part of $\tau_0$.
    In this way they can be easily compared where 
    $\hat U=e^{i\phi_m} U^{\pi/(\phi_M-\phi_m)}$.}
  \label{fig:U_phase_vs_imag_rho}
\end{figure}

In the previous expression we have introduced the following quantities
\begin{align}
\alpha_\RR
=&
2 \sdap \pp_\RR
,
\nonumber\\
e^{\tau_0}
=&
|\alpha_1|^{\alpha_1}\, |\alpha_2|^{\alpha_2}\, |\alpha_3|^{\alpha_3}
,
\nonumber\\
{\cal P}^i
=&
\alpha_1\, p^i_{\sN 2} -\alpha_2\, p^i_{\sN 1}
,
\nonumber\\
N^\RR_m
=&
\frac{1}{\alpha_\RR}\, f_m\left( -\frac{\alpha_{\RR+1}}{\alpha_{\RR}} \right)
,
\nonumber\\
f_m(\gamma)
=&
\frac{1}{m!}\,
(m\gamma-1)\dots (m\gamma-m+1)
,
\nonumber\\
\bar N^\RR_m
=&
\left( e^{\tau_0} \right)^{\frac{m}{\alpha_\RR}}\,
N^\RR_m
.
\label{eq:Mandelstam_N_bar_N_f_etc}
\end{align}
The interaction time $\tau_0$ is the real part of the following map
\begin{align}
  \rho= \tau + i\sigma_M
  =&
  2 \sdap\, \pp_{\sN 1}\, \log(u-1)
  +
  2 \sdap\, \pp_{\sN 2}\, \log(u)
  ,
\end{align}
when evaluated at stationary point
$ \left. \frac{d \rho}{ d u} \right|_{u_0}=0$.
The previous map from the upper half plane to the strip
means that the vertexes are at $x=1, 0, \infty$.
Notice moreover that we have written $\sigma_M$ to signal that this
variable has the usual meaning of worldsheet space coordinate but it
has not the usual range as shown in fig. \eqref{fig:U_phase_vs_imag_rho}.

If we consider the ranges we see that we have to compare
$e^{\oh (\tau+ i \sigma_M)}$ with
$\hat U = (u-x_1)^{\sdap \ukp_{\sN 1}}\, (u-x_2)^{\sdap \ukp_{\sN 2}}\,
(u-x_3)^{\sdap \ukp_{\sN 3}} $
and send $x_3 \rightarrow \infty$ by analytic continuation

In order to compare with our expression we need
\begin{align}
  {\cal P}^i
  =&
  2 \sdap \ukp_{\sN \RR}
  \left( \Alc_{\sN {\RR+1} 0}^{i} - \rho_{\SS \RR} \Alc_{\sN \RR 0}^{i}
  \right)
  ,
  \nonumber\\
  f_m\left( -\frac{\alpha_{\RR+1}}{\alpha_{\RR}} \right)
  =&
  \frac{1}{m}\,
       { -m \rho_{\RR-1\, \RR} -1 \choose m-1}
       ,
\end{align}
when we identify Mandelstam momenta with DDF momenta as
$\pp=\ukp$ and $\sdap p^i_{\sN \RR} = \Alc_{\sN \RR 0}^i$.

Now we can exam the expression given in
eq. \eqref{eq:Reggeon_rewritten_v1}
considering the terms ${\Alc_{\sN \RR n}^i} \Alc_{\sN \SS 0}^i$
and make use of the integrals computed in \cite{Biswas:2024mdu}
\begin{align}
  -
  \sum_{\RR, \SS=1}^3
  \sum_{i=2}^{D-2} \sum_{m=1}^\infty
  &
  \frac{\Alc_{\sN \RR m}^i}{m}\,
  \Alc_{\sN \SS 0}^i\,
  \oint_{z=\xs} \frac{d z}{2\pi i}
  \frac{ \Zlc{ \RR}(z)^ {-m} }{ z -\xs}
  \nonumber\\
  =&
  -
  \sum_{i=2}^{D-2} \sum_{m=1}^\infty
  \sum_{\RR=1}^3
  \frac{\Alc_{\sN \RR m}^i}{n}\,
  \oldI^i_{\sN \RR\, m}
  \nonumber\\
  =&
  -
  \sum_{i=2}^{D-2} \sum_{m=1}^\infty
  \sum_{\RR=1}^3
  \frac{\Alc_{\sN \RR n}^i}{m}\,
  \left(
  \Alc_{\sN {\RR+1} 0}^{i} - \rho_{\SS \RR} \Alc_{\sN \RR  0}^{i}
  \right)\,
  {\hat \oldI}_{\sN \RR\, m,\, \RR+1}
,
\end{align}
where the explicit evaluation gives
\begin{align}
  {\hat \oldI}_{\sN \RR\, m,\, \sN {\RR+1} }
  =&
  -
    \left(
    x_{1 2}^{\sdap \ukp_{\sN 3}}\,
    x_{1 3}^{\sdap \ukp_{\sN 2}}\,
    x_{2 3}^{\sdap \ukp_{\sN 1}}
    \right)^{ \frac{m}{\sdap \ukp_{\sN \RR}} }
      \,
    { - m\rho_{\RR-1\, \RR} -1 \choose m-1}
    .
    \label{eq:comparison_bar_N_Neumann_DDF}
\end{align}

We can now compare with Mandelstam.
If we compare with $\bar N^\RR_m$ we get
\begin{align}
  e^{\oh \tau_0}
  =
  x_{1 2}^{2 \sdap \kp_{\sN 3}}\,
  x_{1 3}^{2 \sdap \kp_{\sN 2}}\,
  x_{2 3}^{2 \sdap \kp_{\sN 1}}\,
  ,
\end{align}
if we compare with $N^\RR_m$ we get
\begin{align}
  1
  =
  x_{1 2}^{2 \sdap \kp_{\sN 3}}\,
  x_{1 3}^{2 \sdap \kp_{\sN 2}}\,
  x_{2 3}^{2 \sdap \kp_{\sN 1}}\,
  .
\end{align}
This seems inconsistent but on shell for $N=3$ and considering the
ghost contribution all $x$ dependence cancels.
This simply means that on shell and for $N=3$ we can effectively drop
the $x$ dependence in the Reggeon.
This is also the same statement we wrote on the equivalence of
$|V_3^{(M)} (\tau_0) \rangle$ and $|V_3^{(M)} (0) \rangle$ on shell.

In doing so the two results matches perfectly.



\printbibliography      

\end{document}